\documentclass[10pt,conference]{IEEEtran}

\usepackage[colorinlistoftodos,prependcaption,textsize=tiny]{todonotes}
\usepackage{lipsum} 
\usepackage{graphics}
\usepackage{rotating}
\usepackage{multirow}
\usepackage{cite}
\usepackage{makecell}
\usepackage{marvosym}
\usepackage{color}
\usepackage{pifont}

\usepackage{pxfonts}

\usepackage{wasysym}

\usepackage{stmaryrd}

\usepackage{subfigure}
\usepackage{tabularx}
\usepackage{adjustbox} 

\usepackage{amssymb}

\usepackage{color}
\usepackage[hyphens]{url}
\usepackage{paralist}
\usepackage[all]{xy}
\usepackage{tabu}

\def\sender{$\hbox{$\bullet$}\kern-1.5pt\hbox{$\rightarrow$}\kern-1.5pt\hbox{$\circ$}$\xspace} 
\def\recipient{$\hbox{$\circ$}\kern-1.5pt\hbox{$\rightarrow$}\kern-1.5pt\hbox{$\bullet$}$\xspace} 
\def\both{$\hbox{$\bullet$}\kern-1.5pt\hbox{$\rightarrow$}\kern-1.5pt\hbox{$\bullet$}$\xspace} 

\def\source{$\hbox{$\bullet$}\kern-0pt\hbox{$\cdots$}$\xspace} 
\def\hop{$\hbox{$\cdots$}\kern-2.5pt\hbox{$\bullet$}\kern-2.5pt\hbox{$\cdots$}$\xspace} 

\newcommand{\yes}{\ding{51}\xspace}
\newcommand{\no}{\ding{55}\xspace}

\newcommand{\connb}{$\multimapdotboth$\xspace} 
\newcommand{\uniform}{$\varoast$\xspace} 
\newcommand{\dynamic}{\ding{107}\xspace} 
\newcommand{\static}{$\varocircle$\xspace} 

\newcommand{\uni}{$\rightarrow$\xspace} 
\newcommand{\bi}{$\leftrightarrow$\xspace} 

\newcommand{\sync}{$\cong$\xspace} 
\newcommand{\async}{$\neq$\xspace} 

\newcommand{\semid}{$\astrosun$\xspace} 
\newcommand{\fullyd}{$\fullmoon$\xspace} 

\newcommand{\timeb}{\clock\xspace} 
\newcommand{\eventb}{$\lightning$\xspace} 

\newcommand{\flats}{$\cdots$\xspace} 
\newcommand{\hierarchical}{\ding{68}\xspace} 

\newcommand{\byuser}{$\smiley$\xspace}
\newcommand{\secrestr}{\Stopsign\xspace}
\newcommand{\all}{\CircledA\xspace}
\newcommand{\netrestr}{\Mundus\xspace}

\newcommand{\fair}{$\equiv$\xspace}

\newcommand{\msgb}{\Letter \xspace} 

\def\clientserver{$\hbox{$\bullet$}\kern-0pt\hbox{$\cdots$}$\kern-2.5pt\hbox{$\bullet$}\xspace} 

\def\p2p{$\hbox{$\bullet$}\kern-0pt\hbox{$\cdots$}\kern-2.5pt\hbox{$\bullet$}\kern-0pt\hbox{$\cdots$}\kern-2.5pt\hbox{$\bullet$}$\xspace} 
\def\hybrid{$\hbox{$\bullet$}\kern-0pt\hbox{$\cdots$}$\kern-2.5pt\hbox{$\circ$}\kern-0pt\hbox{$\cdots$}\kern-2.5pt\hbox{$\bullet$}\xspace} 

\newcommand{\fullyc}{$\boxtimes$\xspace} 
\newcommand{\mostlyc}{$\square$\xspace} 
\newcommand{\partiallyc}{$\sqsubset$\xspace} 

\newcommand{\full}{$\CIRCLE$\xspace} 
\newcommand{\partially}{$\LEFTcircle$\xspace} 

\newcommand{\low}{\textup{L}\xspace} 
\newcommand{\high}{\textup{H}\xspace} 
\newcommand{\midl}{\textup{M}\xspace} 

\usepackage{xspace}
\newcommand{\etal}{\textit{et al.}\xspace}

\newcommand{\ie}{\textit{i.e.,}\xspace}
\newcommand{\eg}{\textit{e.g.,}\xspace}

\mathchardef\-="2D

\renewcommand{\paragraph}[1]{\medskip \noindent \textbf{#1.\ }}

\newlength{\Oldarrayrulewidth}
\newcommand{\Cline}[2]{%
  \noalign{\global\setlength{\Oldarrayrulewidth}{\arrayrulewidth}}%
  \noalign{\global\setlength{\arrayrulewidth}{#1}}\cline{#2}%
  \noalign{\global\setlength{\arrayrulewidth}{\Oldarrayrulewidth}}}

\usepackage{array}
\IEEEoverridecommandlockouts 

\begin{document}

\title{A Survey on Routing in Anonymous Communication Protocols}
%
%

\author{
\IEEEauthorblockN{Fatemeh Shirazi}
\IEEEauthorblockA{KU Leuven \\ ESAT/COSIC and iMinds }
\and
\IEEEauthorblockN{Milivoj Simeonovski}
\IEEEauthorblockA{CISPA, Saarland University}
\IEEEauthorblockA{Saarland Informatics Campus}
\and
\IEEEauthorblockN{Muhammad~Rizwan~Asghar}
\IEEEauthorblockA{Department of Computer Science}
\IEEEauthorblockA{The University of Auckland }\\
\and
\IEEEauthorblockN{}
\and
\makebox[.56\linewidth]{Michael Backes}\\CISPA, Saarland University \& MPI-SWS\\ Saarland Informatics Campus \\
\and
\IEEEauthorblockN{Claudia Diaz}
\IEEEauthorblockA{KU Leuven \\ ESAT/COSIC and iMinds }
\and
\IEEEauthorblockN{     }
}


\maketitle


\begin{abstract}


The Internet has undergone dramatic changes in the past 15 years, and now forms a global communication platform that billions of users rely on for their daily activities. 
While this transformation has brought tremendous benefits to society, it has also created new threats to online privacy, ranging from profiling of users for monetizing personal information to nearly omnipotent governmental surveillance. 
As a result, public interest in systems for anonymous communication has drastically increased.  
Several such systems have been proposed in the literature, each of which offers anonymity guarantees in different scenarios and under different assumptions, reflecting the plurality of approaches for how messages can be anonymously routed to their destination. 
Understanding this space of competing approaches with their different guarantees and assumptions is vital for users to understand the consequences of different design options.

In this work, we survey previous research on designing, developing, and deploying systems for anonymous communication.
To this end, we provide a taxonomy for clustering all prevalently considered approaches (including Mixnets, DC-nets, onion routing, and DHT-based protocols) with respect to their unique routing characteristics, deployability, and performance.
This, in particular, encompasses the topological structure of the underlying network; 
the routing information that has to be made available to the initiator of the conversation; 
the underlying communication model; 
and performance-related indicators such as latency and communication layer.
Our taxonomy and comparative assessment provide important insights about the differences between the existing classes of anonymous communication protocols, and it also helps to clarify the relationship between the routing characteristics of these protocols, and their performance and scalability.


%

\end{abstract}




\section{Introduction}

The Internet has evolved from a mere communication network used by millions of users to a global platform for social networking, communication, education, entertainment, trade, and political activism used by billions of users.
In addition to the indisputable societal benefits of this transformation,
the mass reach of the Internet has created new powerful threats to online privacy. 

The widespread dissemination of personal information that we witness today in social media platforms and applications is certainly a source of concern. 
The disclosure of potentially sensitive data, however, not only happens when people deliberately post content online, but also inadvertently by merely engaging in any sort of online activities. 
This inadvertent data disclosure is particularly worrisome because non-expert end-users cannot be expected to understand the dimensions of the collection taking place and its corresponding privacy implications. 

Widely deployed communication protocols only protect, if at all, the content of conversations, but do not conceal from network observers who is communicating with whom, when, from where, and for how long. 
Network eavesdroppers can silently monitor users' online behavior and build up comprehensive profiles based on the aggregation of user communications' metadata. 
Today, users are constantly tracked, monitored, and profiled, both with the intent of monetizing their personal information through targeted advertisements, and by nearly omnipotent governmental agencies that rely on the mass collection of metadata for conducting dragnet surveillance at a planetary scale.

Anonymous Communication (AC) systems have been proposed as a technical countermeasure to mitigate the threats of communications surveillance. 
The concept of AC systems was introduced by Chaum~\cite{Chaum:1981:mixnets} in 1981, with his proposal for implementing an anonymous email service that aimed at concealing who sent emails to whom. 
The further development of this concept in the last decades has seen it applied to a variety of problems and scenarios, such as anonymous voting~\cite{Sako:1995:Voting,Jakobsson:2002:Voting}, Private Information Retrieval (PIR)~\cite{Dingledine:2000:FreeHaven}, censorship-resistance~\cite{Waldman:2000:Publius, Waldman:2001:Tangler}, anonymous web browsing~\cite{Goldschlag:1996:onionrouting}, hidden web services~\cite{Dingledine:2004:Tor}, and many others. 

Public interest in AC systems has strikingly increased in the last few years. 
This could be explained as a response to recently revealed dragnet surveillance programs, the fact that deployed AC networks seem to become (according to leaked documents\footnote{\url{https://wikileaks.org/}}) a major hurdle for communications surveillance, and to somewhat increased public awareness on the threats to privacy posed by modern information and communication technologies. 

The literature offers a broad variety of proposals for anonymity network designs. 
Several of these designs have been implemented, and some are successfully deployed in the wild. 
Of the deployed systems, the most successful example to date is the Tor network, which is used daily by about two million people~\cite{tormetrics}. 

Existing designs take a variety of approaches to anonymous routing for implementing the AC network.
Routing determines how data is sent through the network, and it as such constitutes the central element of the AC design, determining to a large extent both security and performance of the system.
These approaches rely on different threat models and sets of assumptions, and they provide different guarantees to their users. 
Even though survey articles on AC systems exist~\cite{surveyattacks:2015, Sampigethaya:2006:MixNet, Conrad:2014:I2PTor, AlSabah:2015:Tor-Survey, Ren:2010:Survey, Edman:2009:Survey, Danezis:2008:Survey, Serjantov:2004:Anonymity, Raymond:2000:Survey},
we still lack a systematic understanding, classification, and comparison of the routing characteristics of the plurality of existing AC approaches.

The purpose of this survey is to provide a detailed overview of the routing characteristics of current AC systems, and to examine how their features determine the anonymity guarantees offered by those systems, as well as its overall performance. 
To this end, we first \emph{identify the routing characteristics} that are relevant for AC protocols and provide a \emph{taxonomy} for clustering the  systems with respect to their routing characteristics, deployability, and performance. 
Then, we apply the taxonomy to the extensive scope of existing AC systems, in particular including Mixnets, DC-nets, onion routing systems, and DHT-based protocols. 
Finally, we discuss the relationship between the different routing decisions, and how they affect performance and scalability.



%
%
	%
	%
	%
	%
	%
	%
	%

\paragraph{Outline}
Section~\ref{sec:characteristics} provides our taxonomy for anonymous routing and describes the various routing features and dimensions that we are considering for our evaluation. 
Section \ref{sec:categorization} gives a compact tabular overview describing the classification of existing systems in our taxonomy and 
reviews existing AC systems with respect to their routing characteristics, substantiating the compact overview provided
in the previous section.
Section \ref{sec:discussion} discusses the relationship between routing decisions and security and anonymity goals, and shares some lessons learned.
Section \ref{sec:conclusion} concludes the paper.


\section{Anonymous Routing Protocol Characteristics}
\label{sec:characteristics}

This section first introduces the routing characteristics considered in our taxonomy, and then discusses deployability, and performance metrics for AC networks. 




\subsection{Routing Characteristics}
\label{subsec:routing}

Generally, routing in a communication network refers to the selection of nodes for relaying communication through the network. 
Routing schemes, however, require some essential design components. 
For anonymous communication, we consider four building blocks that are relevant to routing in AC networks. 
These building blocks are node management, transfer/retrieval of node information to/by the routing decision maker, path selection, and forwarding or relaying; 
where path selection is the main design component of routing schemes for AC protocols.   

Several taxonomies and classifications for routing protocols have been proposed in the literature~\cite{Bell:1986:routingreview, Feeney:1999:routingtaxonomy,Zou:2002:routingclassification}. 
However, AC networks aim to conceal the metadata of communications and thus have security requirements that make them fundamentally different from other networks. 

In this section, we present a classification for anonymous routing protocols. 
Our classification (see Tables~\ref{table:mixdcnets} and \ref{table:onionrouting}) is an adaptation from Feeney's taxonomy~\cite{Feeney:1999:routingtaxonomy}, which classifies the routing characteristics of mobile ad hoc networks into four categories: 

\begin{enumerate}

\item \emph{Communication model} describes whether the communication is based on single-channels or multi-channels.  

\item \emph{Structure} describes whether or not nodes are treated equally. 

\item \emph{State information} describes where the topology information is maintained.

\item \emph{Scheduling} describes whether the information about routes is maintained at the source or is instead computed on-demand.

\end{enumerate}

This taxonomy does not address several relevant design features of AC networks, such as probabilistic node selection for constructing circuits, and security considerations for protecting routing information from different network adversaries. 
In addition, not all the characteristics identified by Feeney are relevant to AC routing. 
For example, the distinction between single- and multi-channel features is not relevant in overlay networks, which constitutes a standard design choice for many AC networks. 


We redefine Feeney's criteria to account for design choices that are relevant to anonymous routing protocols. 
We distinguish three groups of features inspired by Feeney's categories: \emph{network structure}, \emph{routing information}, and \emph{communication model}:  

\begin{enumerate}

\item \emph{Network structure} describes the characteristics of the anonymous relays, the connections between them, and the underlying network topology. 

\item \emph{Routing information} describes the network information available to entities deciding on the route of an anonymous connection. 

\item \emph{Communication model} defines the entities that make the routing decisions and describes how these decisions are made.

\end{enumerate}
%
%
In what follows, we describe these features in more details, including their various sub-features and corresponding notation symbols used to denote individual feature instantiations. 
We refer to Table~\ref{table:overview} for a general overview of the resulting taxonomy.

\newcommand{\symbolsep}{$\;$}
\begin{table*}[ht!]
\centering
\caption{Overview of the Protocol Routing Characteristics}
\label{table:overview}
		\extrarowsep=3pt
\begin{tabular}{|m{0.5cm}|c|l|c|c|}
\hline
                                    \multicolumn{3}{|c|}{\textbf{Feature Name}}   & \textbf{Description} & \textbf{Instantiation and Symbols}\\ \hline\hline
\multirow{6}{*}[-1em]{\begin{sideways}\makecell[r]{\textbf{Network} \\ \textbf{Structure}}\end{sideways}}   & \multicolumn{2}{l|}{Network topology}  &  Degree of node connectivity in the network  &   \fullyc (fully) \symbolsep \mostlyc (mostly) \symbolsep \partiallyc (partially)             \\ \cline{2-5} 
                                     & \multirow{2}{*}{\makecell[c]{Connection \\ type}} & Direction    &    Data flow in connections &  \uni (unidirectional) \symbolsep \bi (bidirectional)         \\ \cline{3-5} 
                                     &                                  & Synchronization       &   Timing model for connection establishment and data sending &  \async (asynchronous) \symbolsep \sync (synchronous)            \\ \cline{2-5} 
                                     & \multirow{3}{*}[-0.5em]{Symmetry}        & Roles          &     Users operating as relays & \makecell[c]{\p2p (peer-to-peer) \symbolsep \clientserver (client-server) \symbolsep \\ \hybrid (hybrid) }             \\ \cline{3-5} 
                                     &                                  & Topology          &   Node topology for routing      & \flats (flat) \symbolsep \hierarchical (hierarchical)                    \\ \cline{3-5} 
          &  & Decentralization &      Degree of decentralization for non-routing services &  \semid (semi decentralized) \symbolsep  \fullyd  (fully decentralized)        \\ \hline\hline
\rule{0pt}{1ex}  \multirow{2}{*}[1.5ex]{\begin{sideways}\makecell[c]{\textbf{Routing} \\ \textbf{Info}}\end{sideways}}        & \multicolumn{2}{l|}{Network view}                   &  Network view necessary for making routing decisions       &    \full (complete) \symbolsep \partially (partial)           \\  \cline{2-5} 
                                     & \multicolumn{2}{l|}{Updating}                    &      Triggers for routing information updates           &    \timeb (periodic) \symbolsep \eventb (event-based)         \\ \hline\hline
\multirow{5}{*}[-1em]{\begin{sideways}\makecell[c]{\textbf{Communication} \\ \textbf{Model}}\end{sideways}} & \multicolumn{2}{l|}{Routing type}     &      Node  selection per route     &  \source (source-routed) \symbolsep \hop (hop-by-hop)      \\ \cline{2-5} 
                                     & \multicolumn{2}{l|}{Scheduling}                         &   Prioritization of traffic           &  $\equiv$  (fair) \symbolsep $\Diamonddot$ (prioritized)       \\ \cline{2-5} 
                                     & \multirow{3}{*}[-0.5em]{\makecell[c]{Node \\ selection}}  & Determinism  &  Determinism of node selection  &  \yes (deterministic) \symbolsep   \no (non-deterministic)       \\ \cline{3-5} 
                                     &                                  & Selection set        & 
Permissible set of nodes per route    & \makecell[c]{\all (all)  \symbolsep \secrestr (restricted, security)  \symbolsep  \\ \netrestr  (restricted, network)  \symbolsep   \byuser (user-based)  }         \\ \cline{3-5} 
                                     &                                  & Selection probability   &  Node selection probability per route  & \makecell[c]{ \uniform (uniform) \symbolsep \static (weighted, static) \symbolsep \\ \dynamic (weighted, dynamic)}            \\ \hline\hline
\multirow{4}{*}{\begin{sideways}\makecell[c]{\textbf{Performance, } \\ \textbf{ Deployability}}\end{sideways}} & \multicolumn{2}{l|}{Latency}             & Protocol latency             &\makecell[c]{\low (low-latency) \symbolsep \high (high-latency) \\\midl (mid-latency) }    \\ \cline{2-5} 
                                     & \multicolumn{2}{l|}{Communication mode}                     &   Longevity of connections          &  \connb (connection-based)  \symbolsep \msgb (message-based)            \\ \cline{2-5} 
                                    & \multicolumn{2}{l|}{Implementation}                        &   Implemented    &  \yes (yes)  \symbolsep \no (no)           \\ \cline{2-5} 
& \multicolumn{2}{l|}{Code availability}                     &   Open source         &  \yes (yes)  \symbolsep \no (no)             \\ \cline{2-5} 
\hline

\end{tabular}
\end{table*}


\subsubsection{\textbf{\emph{Network Structure}}} 

We consider first the network features that are relevant to anonymous routing. 
These are, specifically, features relating to: 
(a) the \emph{\textbf{topology}} of the network, which describes how nodes are connected; 
(b) the \emph{\textbf{connection type}}, describing the characteristics of the connections between nodes; and 
(c) \emph{\textbf{symmetry}}, describing whether the entities participating in the network are all similar, or if they can take on different roles and responsibilities for routing data through the network. 

\begin{enumerate}[a)]

\item \textbf{Topology.} 
The topology describes the arrangement of various elements of the network, such as routers and communication links between those routers. 
We only take the logical topology of the network into account, which determines how data flows within it. 
We note that physical topology characteristics, such as the geographical location of computers, sometimes matters in anonymous routing decisions, for example when considering adversaries that control an Autonomous System (AS)~\cite{Feamster:2004:LocationDiversity,Edman:2009:Tor-AS-Aware}. 

We consider the network as a graph in which the routers are represented by graph nodes. 
An edge between two nodes exists if the routing strategy allows those two nodes to be directly connected as part of the same anonymous circuit. 

The connectivity of nodes varies widely across AC network designs, and the advantages and disadvantages of high and low levels of connectivity have been the subject of debate for over a decade~\cite{Boehme:2005:MIXvsP2P}. 

Restricted routing proposals~\cite{Danezis:2003} have shown that for high-latency applications, partially connected networks with certain topological characteristics (\eg based on expander graphs) provide optimal anonymity and latency trade-offs and mitigate certain attacks. 
These results further emphasize the impact of network connectivity features for anonymous routing. 

We classify anonymity networks into three categories according to their connectivity: \emph{fully connected}, \emph{mostly connected}, and \emph{partially connected} networks. 

\begin{itemize}

\item We consider a network to be \emph{fully connected} (\fullyc)\footnote{In parenthesis, we define the symbol or the keyword that is used in the comparative Tables~\ref{table:mixdcnets} and \ref{table:onionrouting} to indicate the corresponding characteristic.} when nodes can potentially connect to most (or all) other nodes (our rule of thumb is that a node on average should be able to connect to at least 95\% of the other nodes; this allows us to include systems that
only exclude a small number of connections in order to prevent certain special cases from occurring). 

\item We call a network \emph{mostly connected} (\mostlyc) if its nodes can potentially connect to at least half the other nodes. 

\item Finally, in \emph{partially connected} (\partiallyc) networks nodes only connect to a relatively small subset of the whole network.

\end{itemize}



Higher connectivity in the network topology leads to better resilience (availability) against node failure, such as Denial of Service (DoS) attacks, such resilience might have in turn a positive influence on anonymity~\cite{Boehme:2005:MIXvsP2P}.

On the other hand, eliminating connections that might induce security problems, such as the connection between two nodes from the same IP family that may be easier to control by an adversary, but can be beneficial to anonymity. 
The same holds for eliminating connections that would induce higher latency, which would, in turn, improve the performance of the system.




\item \textbf{Connection Type.} 
Here, we consider the \emph{direction} and \emph{synchronization} of connections. 
As far as the direction is concerned, we consider the following options:

\begin{itemize}

\item A connection is \emph{unidirectional} (\uni) if the data flow between two entities can only be in one direction. 

\item A connection between two entities is \emph{bidirectional} (\bi) if data can flow in both directions and the same connection is used for sending back the response to a received message. 

\end{itemize}

Typically, interactive applications, such as web browsing, require bidirectional channels, while non-interactive applications, such as email, can just close the connection as soon as the message has been forwarded. 
In the first case, short-lived session keys can be setup to achieve forward secrecy properties; however, in non-interactive applications, such as email, forward secrecy is harder to achieve. 

Bidirectional circuits have the advantage that they induce less overhead in terms of circuit construction. 
Unidirectional connections have the advantage that they are less vulnerable to timing attacks, as a malicious node can only observe data flowing in one direction, which is less informative than bidirectional connections in which patterns of requests and response are visible to all nodes in the path. 
However, note that in unidirectional connection, a larger number of nodes are going to be involved in relaying the communication between a sender and a receiver.




Further, we consider whether the anonymity system involves connection \emph{synchronization}:

\begin{itemize}

\item A connection is \emph{asynchronous} (\async) if the establishment of connections and relaying of messages is initiated by a user without any timing coordination with other participants. 

\item Connections are \emph{synchronous} (\sync) if they begin and end at specific timings and messages are also relayed at specific moments in time, based on some timing coordination between network entities. 

\end{itemize}

Asynchronous systems are conceptually simpler as they impose fewer constraints on the activity of network participants. 
However, the distinct timing of actions leaks information valuable to perform traffic analysis and, for example, reveals long-term communication patterns~\cite{Danezis:2003:disclosure} or perform end-to-end correlation attacks~\cite{Levine:2004:Mix-Attack,Bauer:2007:Attack, Zhu:2010:CorrelationAnon}. 

Synchronous systems are often more difficult to engineer and come with a performance or usability penalty; moreover, secure and reliable time becomes an additional dependency of the system, and a possible point of failure or vulnerability to attack. 
However, synchronization constitutes a very powerful design feature to offer robust anonymity guarantees in the presence of powerful adversaries 
because it disables trivial end-to-end correlation attacks based on start and end times of connections~\cite{Murdoch:2007:Traffic-Analysis}, and other timing data that synchronization makes less granular, enabling the aggregation of participants, connections, and events in \emph{anonymity sets}. 
Synchronous anonymity systems were proposed in the early 1990s by Pfitzmann \etal to anonymize ISDN telephony calls~\cite{Pfitzmann:1991:ISDN-MIXes}. 
These proposals were both feasible from an engineering perspective (compatible with the network requirements and introducing a low-efficiency cost), and clearly spelled-out anonymity guarantees as well as full unobservability for local calls. 



\item \textbf{Symmetry.} 
We consider symmetry in the roles of the network entities. 
An anonymity system is intuitively ``more symmetric'' when all the participating entities have similar roles and responsibilities, and ``less symmetric'' if there are different roles, capabilities, and trust assumptions among the entities that participate in the routing.  

We thus first examine the overlap between the \emph{roles} of end-users who initiate communications and relaying nodes. 
We distinguish three types of systems. 

\begin{itemize}

\item We classify a system as \emph{peer-to-peer} (\p2p), when end-users are expected (often even obliged) to operate as relaying nodes in order to use the AC network. 

\item At the other end of the spectrum, in \emph{client-server} (\clientserver) systems, users are not expected (often even forbidden) to operate as relaying nodes on order to use the system. 

\item We call a system \emph{hybrid} (\hybrid) if it combines characteristics of both \emph{peer-to-peer} and \emph{client-server} systems, \ie end-users may or may not operate as relaying nodes. 

\end{itemize}

These different levels of symmetry come with advantages and disadvantages~\cite{Boehme:2005:MIXvsP2P}. 
Peer-to-peer systems can better scale as the number of users grows, because new users also increase the capacity of the network. 
Further, peer-to-peer networks are more resilient to node failures and have better availability properties. 
In client-server architectures, however, it is possible to run nodes more reliably and securely (as nodes are not necessarily run by laymen end-users), which in particular helps in handling liability issues with respect to complaints. 
Having end users run just client software has a lower cost for end-users in terms of resources, and offers opportunities for simpler, and thus often more usable, client software.

Second, we distinguish whether nodes are organized in a flat or a hierarchical structure with respect to routing. 
We call the resulting feature the \emph{topology}:

\begin{itemize}

\item  A network has a \emph{flat} (\flats) structure if every node has the same importance and rank when making routing decisions. 

\item A network has a \emph{hierarchical} (\hierarchical) structure if nodes have different capabilities and priorities towards the routing algorithm. 

\end{itemize}

Hierarchical structures are often introduced to improve efficiency and performance. 
However, a non-flat hierarchy can make the network less resilient to attacks, as the failure of a node that is placed high in the hierarchy has a severe impact on the performance of the network.  

The third and last dimension of symmetry addresses the degree of \emph{decentralization} of network services other than (but auxiliary to) the routing itself. 
Note that we are not considering \emph{centralized} models because they are a single point of failure for surveillance and insecure by design.

\begin{itemize}

\item A network is \emph{semi decentralized} (\semid) if it includes one or a small number of entities performing a service critical to routing (\eg compiling and distributing network directory information). 
This accounts for the fact that especially high levels of trust placed on these entities, which constitute more of a point of failure than a simple relay.

\item A network is \emph{fully decentralized} (\fullyd) if the system design does not include entities that have to be especially trusted for the provision of functionalities that enable the routing. 
Fully decentralized systems have a better distribution of trust.

\end{itemize}



\end{enumerate}


\subsubsection{\textbf{Routing Information}} 
We now consider the information available to the entity (or entities) that decides on the route of a connection, and how that information is made available.

\begin{enumerate}[a)]

%

\item \textbf{Network View}. 
This determines the completeness of information available to establish a route.

\begin{itemize}

\item The routing decision-maker has a \emph{complete view} (\full) of the system if routing information about all nodes is available to her. 

\item The decision maker has a \emph{partial view} (\partially) of the system if the routing information available to her only covers a subset of the nodes that form the AC network. 

\end{itemize}

A complete view allows the decision maker to choose among the full set of nodes. 
However, a partial view improves the scalability of the network, as the distribution of routing information for the full network may consume significant bandwidth and network resources. 
There are also some attacks that become possible when the routing decision makers only have a partial view of the network. 
For example, route fingerprinting attacks \cite{Danezis:2006:Fingerprinting, Danezis:2008:RouteBridgingAttacks} are possible if each user knows different subsets of routers. 
In these attacks, the initiator of a connection can be identified by the nodes that make up the route, since typically a very small number of users will know a certain combination of network nodes. 



\item \textbf{Updating}. 
This determines how frequently routing information is updated. 

\begin{itemize}

\item Routing information is updated \emph{periodically} (\timeb) if it is updated in predefined time intervals.

\item Routing information is updated \emph{event-based} (\eventb) if the updates are triggered by events in the network other than timeouts.

\item No updating mechanism is in place (\no).

\end{itemize}

\end{enumerate}

\subsubsection{\textbf{\emph{Communication Model}}}
\label{communication-model} 
We finally consider features that describe the creation of anonymous routes. 
\begin{enumerate}[a)]

\item{\textbf{Routing Type.}} 
This refers to the selection of nodes to determine a route.

\begin{itemize}

\item The routing decision is \emph{source-routed} (\source) if the initiator of the communication selects the set of nodes that will form the anonymous route. 

\item The routing decision is \emph{hop-by-hop} (\hop) (also called ``random routing'') if the initiator only selects the first relay node, which in turn picks the second, and so on, until the message reaches its final destination. 

\end{itemize}

Source-routing enables the initiator to pick nodes she trusts, and prevents adversaries from biasing the node selection towards compromised nodes. 
A variation of the basic source-routed model is found in some systems that provide receiver anonymity. 
In these systems, the initiator and the receiver select, respectively, the first and second halves of the route, which are joined in the middle at a rendezvous point. 
An advantage of hop-by-hop routing is that even if the initiator only knows a subset of nodes, her connections might be routed throughout the whole network, mitigating route fingerprinting attacks~\cite{Danezis:2006:Fingerprinting}. 
In literature, other node selection strategies have been proposed, which we have not taken into consideration such as dynamic routing schemes using distance vector routing (\ie \cite{Perkins:2003:RFC3561}) and link-state routing (\ie \cite{Moy:1998:RFC2328}). 
Such algorithms are often disregarded for AC networks because of the predictability they offer, which is in conflict with anonymity. 

\item \textbf{Scheduling.} 
This refers to the way a node serves incoming scheduling requests.

\begin{itemize}
\item \emph{Fair} ($\equiv$) scheduling means that all types of connection are treated same. 

\item \emph{Prioritized} ($\Diamonddot$) scheduling means that certain connections are given priority over others.

\end{itemize}

Prioritized scheduling can improve performance and reduce congestion. 
However, differential treatment of traffic may undermine anonymity as the traffic of different priorities would be distinguishable and thus not conform a single (larger) anonymity set. 
An example of prioritized scheduling is when the scheduling follows an economic model, which might mitigate flooding attacks~\cite{Grothoff:2003:EconomicModelGnuNet}.

\item \textbf{Node Selection.} 
This refers to the protocol features that determine which nodes are selected to be part of an anonymous route. 
The number of nodes that are selected to form the anonymous connection can either be fixed (deterministically) or be computed probabilistically according to some distribution.

\begin{itemize}

\item Node selection can either be \emph{deterministic} (\yes) or non-deterministic (probabilistic) (\no).

\end{itemize}
To characterize node selection, we consider the \emph{selection set} that determines which nodes are eligible for being on the route, and the \emph{selection (probability) distribution} that describes the likelihood of each of the nodes in the selection set being chosen for a route. 


\begin{itemize}

\item  The selection set may contain \emph{all nodes} (\all) of the network.

\item It may contain a \emph{security-restricted subset} (\secrestr) of all network nodes, \ie a subset that is selected according to some \emph{security-restrictions}, for example establishing that all the nodes in a route must be in different /16 IP subnets.

\item  It may contain a \emph{network-restricted subset} (\netrestr) of all network nodes, \eg a subset aimed at guaranteeing the quality of the communication, by for example avoiding congested links and nodes.

\item And finally, the selection set may be user-specific, considering \emph{user preferences and trust assumptions} (\byuser).
 

\end{itemize}

We are left to define the selection probability with which individual nodes are chosen.

\begin{itemize}

\item The probability distribution that describes how nodes are selected may be \emph{uniform} (\uniform).

\item The probability distribution is \emph{statically weighted}, \ie weighted based on \emph{general, static  parameters} (\static), for example the bandwidth of the nodes.

\item The probability distribution is \emph{dynamically weighted} based on \emph{state-specific dependencies} (\dynamic), for example the nodes' response time.

\end{itemize}


Even for general parameters, weighted selection often requires frequent updates so they reflect the current state of the network. 
In other words, we consider parameters that are calculated in real-time to be \emph{dynamic} biases, and parameters based on routing information that is unchanged until the next periodic update to be \emph{static}. 
Uniform selection typically offers better anonymity levels, while weighted selection often improves performance.

\end{enumerate}

%


\subsection{Performance and Deployability} 
In addition to the routing characteristics identified before, we finally identify the following list of metrics that can be used to evaluate performance and deployability characteristics of AC protocols.

\begin{enumerate}

 \item {\textbf{Latency.}} 
 In the literature, AC protocols are usually classified into two performance categories: 

\begin{itemize}

\item Protocols with \emph{low-latency} (\low) incorporate no latency to the communication and typically support applications that require real-time communication (\eg web browsing). 

\item Protocols with \emph{high latency} (\high) do not require real-time communications and support applications that can tolerate a certain delay between requests and responses (\eg email communication). 

\item Protocols with \emph{mid latency} (\midl) introduce a random delay and may induce a restricted latency; hence, these protocols support applications that can tolerate a restricted delay between requests and responses (\eg file sharing). 

\end{itemize}


	
 
	
\item {\textbf{Communication Mode.}} 
 We distinguish two kinds of communication modes, depending on the longevity of individual connections.
 
\begin{itemize}

\item We classify protocols as \emph{connection-based} (\connb) if routes between senders and receivers are maintained for a certain amount of time and used for exchanging multiple data transfers.

\item If routes are created just to send a message and no state is maintained for further exchanges, then we classify a protocol as \emph{message-based} (\msgb).

\end{itemize}


\item {\textbf{Implementation and Code Availability.}} 
This indicates whether or not a prototype of the protocol has been implemented, and if the code is publicly available, respectively. 
In both cases, the answer is either yes (\yes) or no (\no).
 
%



\end{enumerate}




\section{Routing Classification of AC Protocols}
\label{sec:categorization}

In this section, we present a categorization of AC protocols. 
We have classified these protocols into four main families: 
(1) Mixnet-based protocols, 
(2) Onion Routing-based protocols, 
(3) Random Walk and Distributed Hash Table (DHT)-based protocols, and 
(4) DCNet-based protocols
(5) Miscellaneous, containing a few protocols that do not fit into the aforementioned categories. 
A few protocols are presented in the most representative category, albeit they can technically fall under other categories as well, \eg Octopus and Torsk are DHT-based, but they also use onion routing.
We summarize our classification of the routing aspects in two comparative tables (namely Table~\ref{table:mixdcnets} and Table~\ref{table:onionrouting}).

\begin{sidewaystable*}[htbp]
     \centering
		\scriptsize

     \caption{Routing Classification of Anonymous Communication Protocols: Mixnet and Onion Routing Protocols}
		\extrarowsep=2pt
       \begin{tabu}{cc|c|[1.5pt] c|c|c|c|c|c|[1.5pt] c|c|[1.5pt] c|c|c|c|c|[1.5pt] c|c|c|c| [1.5pt] c}
\tabucline[1pt]{4-20}
   \multicolumn{1}{c}{} &
         \multicolumn{1}{c}{} & 
          &
         \multicolumn{6}{c|[1.5pt] }{\textbf{Network Structure}} &
         \multicolumn{2}{c|[1.5pt] }{\makecell[c]{\textbf{Routing} \\ \textbf{Information}}} &
         \multicolumn{5}{c|[1.5pt] }{\textbf{Communication Model}} &
         \multicolumn{4}{c|[1.5pt] }{\textbf{Performance and Deployability}}
         \\ 
\tabucline[1pt]{4-20}
      \multicolumn{1}{c}{} & \multicolumn{1}{c}{} & & &
         \multicolumn{2}{c|}{Connection Type} &
         \multicolumn{3}{c|[1.5pt] }{Symmetry} &
         & & & &
         \multicolumn{3}{c|[1.5pt] }{Node Selection} &
          & & & &
         \\ 
\cline{5-9} \cline{14-16}       
					\multicolumn{1}{c}{} &
          \multicolumn{1}{c}{} &
           &
         \begin{sideways}Topology \end{sideways}&
         \begin{sideways}\makecell[c]{Direction} \end{sideways} &
         \begin{sideways}Synchronization \, \end{sideways}&
         \begin{sideways}\makecell[l]{Roles}\end{sideways}&
         \begin{sideways}Hierarchy \end{sideways} &
         \begin{sideways}\makecell[l]{Decentralization} \end{sideways}&
         \begin{sideways}Network view \end{sideways}&
         \begin{sideways}Updating \end{sideways}&
         \begin{sideways}Routing type\end{sideways} &
         \begin{sideways}Scheduling \end{sideways} &
         \begin{sideways}Determinism \end{sideways} &
         \begin{sideways}Selection set \end{sideways} &
         \begin{sideways}\makecell[l]{Selection \\ probability} \end{sideways} &
         \begin{sideways}Latency\end{sideways} &
         \begin{sideways}\makecell[l]{Communication \\ mode}\end{sideways} &
         \begin{sideways}Implementation\end{sideways} &
         \begin{sideways}Code availability\end{sideways}
				\\
         \cline{1-20}

\multicolumn{1}{|c|}{\multirow{12}[6]{*}{\begin{sideways}\textbf{Mixnet-based Protocols}\end{sideways}}} 		
                                    &Chaum's Mix Cascades & \cite{Chaum:1981:mixnets} 								&\fullyc	&\uni	&\async	&\clientserver	&\flats	&\no		&\full 		&\no 	&\source 	&$\equiv$ 	&\yes	&\all			&\static	&\high		&\msgb	&\no	&\no	\\
\cline{2-20} 
\multicolumn{1}{|c|}{}&ISDN and Real-time  & \cite{Pfitzmann:1991:ISDN-MIXes}\cite{Jerichow:1998:realtimemix} 					&\partiallyc 	&\bi 	&\sync 	&\clientserver	&\flats	&\semid		&\partially 	&\no	&\source 	&$\equiv$ 	&\yes 	&\all 			&\static 	&\low 		&\connb &\no 	&\no	\\
\cline{2-20} 
\multicolumn{1}{|c|}{}&Babel &	\cite{Babel:1996} 												&\fullyc 	&\uni	&\async	&\clientserver 	&\flats	&\semid		&\full 		&\no  	&\source/\hop	&$\equiv$	&\no 	&\all			&\uniform	&\high		&\msgb	&\no	&\no	\\
\cline{2-20} 
\multicolumn{1}{|c|}{}&Stop-and-Go-MIXes &\cite{Kesdogan:1998:stopandgo} 									&\fullyc 	&\uni 	&\sync 	&\clientserver 	&\flats	&\semid	 	&\full		&\timeb	&\source	&$\equiv$	&\no	&\all 			&\uniform	&\high		&\msgb	&\no	&\no	\\
\cline{2-20} 
\multicolumn{1}{|c|}{}& Webmixes& \cite{Berthold:2000:JAP,Berthold:2000:Anonymity} 								&\partiallyc	&\bi 	&\sync 	&\clientserver	&\flats &\semid 	&\full 		&\timeb	&\source	&$\equiv$	&\yes 	&\byuser 		&\static 	&\low 		&\connb &\yes 	&\yes 	\\
\cline{2-20} 
\multicolumn{1}{|c|}{}&\makecell[c]{A Reputation System for Mixnets}&\cite{Dingledine:2001:Reliability}						&\fullyc	&\uni	&\sync 	&\clientserver	&\flats	&\semid		&\full		&\timeb &\source/\hop	&$\equiv$	&\no	&\secrestr 		&\dynamic 	&\high 		&\msgb	&\no	&\no 	\\
\cline{2-20} 
\multicolumn{1}{|c|}{}& Reliable Mix cascades &\cite{Dingledine:2003:Reliable-Mix}								&\fullyc	&\uni	&\sync	&\clientserver	&\flats	&\semid		&\full		&\timeb	&\source	&$\equiv$	&\no 	&\all			&\dynamic	&\high		&\msgb	&\no	&\no 	\\
\cline{2-20}       
\multicolumn{1}{|c|}{}& Mixmaster& \cite{moller:2003:mixmaster}											&\fullyc	&\uni	&\async	&\clientserver 	&\flats &\semid		&\partially	&\no  	&\source	&$\equiv$ 	&\no 	&\all			&\uniform 	&\high		&\msgb	&\yes	&\yes 	\\
\cline{2-20} 
\multicolumn{1}{|c|}{}& Mixminion &\cite{Danezis:2003:mixminion} 										&\fullyc 	&\uni	&\async &\clientserver	&\flats &\semid		&\full		&\timeb &\source	&$\equiv$ 	&\no	&\all 			&\uniform 	&\high 		&\msgb	&\yes 	&\yes 	\\
\cline{2-20} 
\multicolumn{1}{|c|}{}&\makecell[c]{Mix-Networks with \\Restricted Routes} &\cite{Danezis:2003} 						&\partiallyc 	&\uni 	&\async &\clientserver 	&\flats &\semid		&\partially	&\timeb &\source	&$\equiv$	&\no 	&\netrestr 		&\static 	&\high		&\msgb	&\no	&\no	\\
\cline{1-20} 
\multicolumn{1}{|c|}{\multirow{11}[6]{*}{\begin{sideways}\textbf{Onion Routing Protocols}\end{sideways}}} 		
			    &Tor & \cite{Dingledine:2004:Tor} 											&\mostlyc &    \bi    & \async 	&\hybrid    	&\flats &\semid    	&\full 		&\timeb &\source 	&$\equiv$ 	&\no 	& \netrestr \secrestr 	& \static & \low &    \connb & \yes & \yes \\
\cline{2-20} 
\multicolumn{1}{|c|}{}&AS-Aware Node Selection& \cite{Edman:2009:Tor-AS-Aware} 									&\mostlyc &    \bi    & \async 	&\hybrid    	&\flats &\semid    	&\full 		&\timeb &\source 	&$\equiv$ 	&\no 	& \netrestr \secrestr 	& \static & \low &    \connb & \yes & \yes \\
\cline{2-20}
\multicolumn{1}{|c|}{}&LASTor & \cite{Akhoondi:2012:LASTor} 											&\mostlyc &    \bi    & \async 	&\hybrid    	&\flats &\semid    	&\full 		&\timeb &\source 	&$\equiv$ 	&\no 	& \netrestr \secrestr 	& \static & \low & \connb & \yes & \no \\
\cline{2-20}
\multicolumn{1}{|c|}{}&Coordinate Node Selection&\cite{Sherr:2009:Tor-Link-BasedSelection} 							&\mostlyc &    \bi    & \async 	&\hybrid    	&\flats & \semid    	&\full 		&\timeb &\source 	&$\equiv$ 	&\no 	& \netrestr \secrestr 	& \static & \low &    \connb & \yes & \yes \\
\cline{2-20} 
\multicolumn{1}{|c|}{}&Tuneup for Node Selection& \cite{Snader:2011:Tor-tuneup} 								&\mostlyc &    \bi    & \async 	&\hybrid    	&\flats &\semid    	&\full 		&\timeb &\source 	&$\equiv$ 	&\no 	& \netrestr \secrestr 	& \makecell[c]{\dynamic} & \low &    \connb & \yes & \no \\
\cline{2-20} 
\multicolumn{1}{|c|}{}&\makecell[c]{Congestion-aware \\ Node Selection} & \cite{Wang:2012:Tor-CongestionAware} 					&\mostlyc &    \bi    & \async 	&\hybrid    	&\flats &\semid    	&\full 		&\timeb &\source 	&$\equiv$ 	&\no 	& \netrestr \secrestr 	& \dynamic & \low &    \connb & \yes & \no\\
 \cline{2-20} 
\multicolumn{1}{|c|}{}&Panchenko Node Selection and Mator& \cite{Panchenko:2012:ImprovingTorPerformance}\cite{Backes:2014:Mator} 		&\mostlyc &    \bi    & \async 	&\hybrid    	&\flats &\semid    	&\full 		&\timeb &\source 	&$\equiv$ 	&\no 	& \netrestr \secrestr 	& \static & \low &    \connb & \yes & \no\\
\cline{2-20} 
\multicolumn{1}{|c|}{}&Torchestra, PCTCP, IMUX & \cite{Gopal:2012:Torchestra}\cite{AlSabah:2013:PCTCP}\cite{Geddes:2014:Imux}			&\mostlyc &    \bi    &\async 	&\hybrid    	&\flats &\semid    	&\full 		&\timeb &\source 	&$\equiv$ 	&\no 	& \netrestr \secrestr 	& \static & \low &    \connb & \yes & \no \\
\cline{2-20} 
\multicolumn{1}{|c|}{}&Conflux& \cite{AlSabah:2012:Conflux} 											&\mostlyc &    \bi    &\async 	&\hybrid    	&\flats &\semid    	&\full 		&\timeb &\source 	&$\equiv$ 	&\no 	& \netrestr \secrestr 	& \dynamic & \low &    \connb & \yes & \no \\
\cline{2-20} 
\multicolumn{1}{|c|}{}&Prioritized Scheduling/ DiffTor & \cite{Tang:2010:Tor-CircuitScheduling} \cite{AlSabah:2012:DiffTor} 			&\mostlyc &    \bi    &\async 	&\hybrid    	&\flats &\semid    	&\full 		&\timeb &\source 	&$\Diamonddot$ 	&\no 	& \netrestr \secrestr 	& \static & \low &    \connb & \yes & \no\\
\cline{2-20} 
\multicolumn{1}{|c|}{}&PIR-Tor & \cite{Mittal:2011:PIR-Tor} 											&\mostlyc &    \bi    &\async 	&\hybrid    	&\flats &\semid    	&\partially 	&\timeb &\source 	&$\equiv$ 	&\no 	& \netrestr \secrestr 	& \static & \low &    \connb & \yes & \no \\
\cline{1-20} 

\Cline{5pt}{1-20}

   \end{tabu}%
     \label{table:mixdcnets}%

 \end{sidewaystable*}

   \begin{sidewaystable*}[htbp]
     \centering
		\scriptsize
         
             \caption{Routing Classification of Anonymous Communication Protocols: DHT-based, DCNets, and Miscellaneous Protocols}
								\extrarowsep=2pt
             \begin{tabu}{cc|c|[1.5pt] c|c|c|c|c|c|[1.5pt] c|c|[1.5pt] c|c|c|c|c|[1.5pt] c|c|c|c| [1.5pt] c}
                
                 \tabucline[1pt]{4-20}
                 \multicolumn{1}{r}{} &
                 \multicolumn{1}{r}{} &
                 &
                 \multicolumn{6}{c|[1.5pt] }{\textbf{Network Structure}} &
                 \multicolumn{2}{c|[1.5pt] }{\makecell[c]{\textbf{Routing} \\ \textbf{Information}}} &
                 \multicolumn{5}{c|[1.5pt] }{\textbf{Communication Model}} &
                 \multicolumn{4}{c|[1.5pt]}{\textbf{Performance and Deployability}}
                 \\
                 \tabucline[1pt]{4-20}
                 \multicolumn{1}{r}{} & \multicolumn{1}{r}{} & & &
                 \multicolumn{2}{c|}{Connection Type} &
                 \multicolumn{3}{c|[1.5pt] }{Symmetry} &
                 & & & &
                 \multicolumn{3}{c|[1.5pt] }{Node Selection} &
                 & & & &
                 \\
                 \cline{5-9} \cline{14-16}      
                 \multicolumn{1}{r}{} &
                 \multicolumn{1}{r}{} &
                 &
                 \begin{sideways}Topology \end{sideways}&
                 \begin{sideways}\makecell[c]{Direction} \end{sideways} &
                 \begin{sideways}Synchronization \, \end{sideways}&
                 \begin{sideways}\makecell[l]{Roles}\end{sideways}&
                 \begin{sideways}Hierarchy \end{sideways} &
                 \begin{sideways}\makecell[l]{Decentralization} \end{sideways}&
                 \begin{sideways}Network view \end{sideways}&
                 \begin{sideways}Updating \end{sideways}&
                 \begin{sideways}Routing type\end{sideways} &
                 \begin{sideways}Scheduling \end{sideways} &
                 \begin{sideways}Determinism \end{sideways} &
                 \begin{sideways}Selection set \end{sideways} &
                 \begin{sideways}\makecell[c]{Selection \\ probability} \end{sideways} &
                 \begin{sideways}Latency\end{sideways} &
                 \begin{sideways}\makecell[l]{Communication \\ mode}\end{sideways} &
                 \begin{sideways}Implementation\end{sideways} &
                 \begin{sideways}Code availability\end{sideways}
                 \\
                 \cline{1-20}

\multicolumn{1}{|c|}{\multirow{10}[4]{*}{\begin{sideways}\makecell[c]{\textbf{DHT-based Protocols}}\end{sideways}}}
		       &Crowds     &    \cite{Reiter:1998:Crowds} 												&\fullyc 	&\bi 	&\async &\p2p 		&\flats 	&\semid 	&\full 		&\eventb 	&\hop 		&$\equiv$ 	&\no 		&\all 			&\uniform  	&\low 	&\msgb 	&\yes 	&\no 	\\
\cline{2-20}                       
\multicolumn{1}{|c|}{} &MorphMix & \cite{Rennhard:2002:MorphMix,Rennhard:2004:MorphMix} 									&\partiallyc	&\bi	&\async &\p2p 		&\flats 	&\semid 	&\partially 	&\eventb 	&\hop 		&$\equiv$ 	&\no 		&\netrestr		&\dynamic 	&\low 	&\connb &\yes 	&\yes 	\\
\cline{2-20} 
\multicolumn{1}{|c|}{}&Torsk & \cite{McLachlan:2009:Torsk} 													&\partiallyc 	&\bi 	&\async &\hybrid	&\flats 	&\semid 	&\partially 	&\eventb    	&\source 	&\fair		&\no 		&\netrestr 		&\uniform 	&\low 	&\connb &\yes 	&\no	\\
\cline{2-20}
\multicolumn{1}{|c|}{} &NISAN &    \cite{Panchenko:2009:NISAN} 													&\partiallyc 	&\bi 	&\async &\p2p 		&\flats 	&\fullyd 	&\partially 	&\eventb  	&\source    	&\fair		&\no 		&\all 			&\uniform 	&\low 	&\msgb 	&\yes 	&\no 	\\%
\cline{2-20} 
\multicolumn{1}{|c|}{} &AP3 & \cite{Mislove:2004:AP3} 														&\partiallyc 	&\bi 	&\async &\p2p 		&\flats 	&\fullyd 	&\partially 	&\eventb    	&\hop 		&\fair		&\no     	&\all 			&\uniform 	&\low   &\connb &\no 	&\no 	\\%
\cline{2-20} 
\multicolumn{1}{|c|}{} &Salsa & \cite{Nambiar:2006:Salsa} 													&\partiallyc 	&\bi	&\async &\p2p or \hybrid&\flats 	&\fullyd 	&\partially 	&\eventb 	&\source    	&\fair		&\no    	&\all 			&\uniform 	&\low 	&\connb &\yes 	&\no   	\\
\cline{2-20} 
\multicolumn{1}{|c|}{} &Octopus & \cite{Wang:2012:Octopus} 													&\partiallyc 	&\bi 	&\async &\p2p 		&\flats 	&\semid 	&\partially 	&\eventb 	&\source 	&\fair		&\no    	&\all 			&\uniform 	&\midl 	&\msgb 	&\yes 	&\no 	\\
\cline{2-20}
\multicolumn{1}{|c|}{} &Freenet Opennet &    \cite{Clarke:2001:FN}    												&\partiallyc 	&\bi    &\async &\p2p   	&\flats 	&\semid 	&\partially 	&\timeb    	&\hop 		&\fair 		&\yes 		&\all 			&\dynamic    	&\low	&\msgb 	&\yes 	&\yes 	\\
\cline{2-20} 
\multicolumn{1}{|c|}{} &Freenet Darknet & \cite{Clarke:2010:FN} 												&\partiallyc 	&\bi 	&\async &\p2p   	&\flats 	&\fullyd 	&\partially 	&\timeb 	&\hop    	&\fair 		&\yes    	&\byuser 		&\dynamic    	&\low	&\msgb  &\yes 	&\yes 	\\
\cline{2-20} 
\multicolumn{1}{|c|}{} &GAP from GNUnet & \cite{Bennett:2002:GN} \cite{Bennett:2003:GAP} 									&\partiallyc	&\bi	&\async	&\p2p		&\flats 	&\fullyd	&\partially 	&\eventb	&\hop		&$\Diamonddot$	&\yes		&\all			&\dynamic	&\midl	&\msgb	&\yes	&\yes	\\
\cline{1-20}  
\multicolumn{1}{|c|}{\multirow{4}[4]{*}{\begin{sideways}\textbf{DCNets}\end{sideways}}} 
&Chaum's DCNet, Revisited DCnet&\cite{Chaum:1988:DCNets}\cite{Waidner:1990:DC}\cite{Golle:2004:DC}			&\fullyc 	&\uni 	&\async &\p2p 		& \flats 	&\no	 	&\full 		&\eventb 	&\source 	&$\equiv$ 	&\yes 		&\all 			&\static	&\high 	&\msgb 	&\no 	&\no	\\
\cline{2-20} 
\cline{2-20} 	
\multicolumn{1}{|c|}{} &Herbivore &	\cite{Goel:2003:Herbivore} 												&\partiallyc	&\uni	&\async	&\p2p 		&\hierarchical	&\semid 	&\partially	&\eventb	&\source	&$\equiv$	&\yes		&\netrestr 		&\static	&\high 	&\msgb 	&\yes 	&\no	\\
\cline{2-20} 
\multicolumn{1}{|c|}{} &Dissent &	\cite{Corrigan-Gibbs:2010:Dissent} 											&\fullyc 	&\uni 	&\async &\p2p 		&\flats 	&\semid 	&\full 		&\eventb 	&\source 	&$\equiv$ 	&\yes 		&\all 			&\static 	&\high 	&\msgb 	&\yes 	&\yes	\\
\cline{2-20} 
\multicolumn{1}{|c|}{} &Dissent in Numbers & \cite{Wolinsky:2012:Dissent, Wolinsky:2012:Dissent-anytrust} 							&\partiallyc 	&\uni 	&\async &\clientserver 	&\hierarchical	&\semid 	&\partially 	&\eventb 	&\source 	&$\equiv$ 	&\yes 		&\netrestr 		&\static 	&\high 	&\msgb 	&\yes 	&\yes 	\\
\Cline{5pt}{1-20}

\multicolumn{1}{|c|}{\multirow{2}[4]{*}{\begin{sideways}\makecell[c]{\textbf{Misc.}}\end{sideways}}}
 &\makecell[c]{Tarzan}	&\cite{IPTPS:2002:Tarzan,Freedman:2002:Tarzan}												&\mostlyc	&\bi 	&\async	&\p2p 		&\flats		&\fullyd	&\full 		&\eventb 	&\source 	&\fair 		&\no 		&\secrestr 		&\uniform	&\low	&\connb &\yes 	&\yes	\\
\cline{2-20} 
\multicolumn{1}{|c|}{} &I2P    & \cite{Web:2014:I2P} \cite{Schimmer:2009:I2P} 											&\mostlyc	&\uni 	&\async &\p2p 		&\flats 	&\fullyd 	&\full 		&\timeb 	&\source     	& $\Diamonddot$	&\no 		&\netrestr \secrestr 	&\dynamic 	&\low 	&\connb &\yes   &\yes 	\\
\cline{2-20} 
\cline{1-20}  
	

   \end{tabu}%
     \label{table:onionrouting}%

\end{sidewaystable*}







We now discuss the AC protocols individually, starting with Mixnet-based protocols 
(from Section~\ref{sec:mixnets} to Section~\ref{sec:remailers}), and then proceeding with 
Onion Routing-based protocols (Section~\ref{sec:onionrouting} and Section~\ref{sec:onionroutingbased}),
DHT-based protocols (Sections~\ref{sec:dht}), DCNet-based protocols (Section~\ref{sec:DCNets}), and finally the class of miscellaneous protocols (Section~\ref{sec:misc}).



\subsection{Mixes}
\label{sec:mixnets}

The idea of anonymous communication was originally proposed by David Chaum in 1981~\cite{Chaum:1981:mixnets} and initiated a new field of privacy research. 
The central concept proposed by Chaum is the use of \emph{mix nodes}, or \emph{mixes} in short.
Mix nodes cryptographically transform messages so that they cannot be traced based on their content. 
Further, mixes shuffle (``mix'') input messages and output them in a reshuffled form. 
Thereby, they hide the input-output relation between individual messages, such that an adversary is not able to establish a correlation between input and output messages.
In Chaumian mixes, the mix node does not output the messages immediately upon arrival, but instead collects a certain number of messages (up to a threshold) into a so-called \emph{batch}, which introduces a delay in message transmission.
The mix shuffles input messages within a batch and flushes them out ordered lexicographically.

\subsection{Mix Selection Strategies}
\begin{figure}[htp!]
\centering
\includegraphics[width=.36\textwidth]{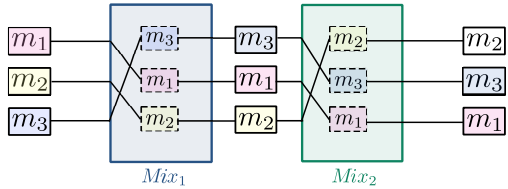}
\caption{A mix cascade with two mixes}
\label{fig:mix}
\end{figure}

In order to distribute trust, Chaum proposed to relay messages through a fixed sequence of mix nodes\footnote{In the literature, a sequence of mixes is usually referred to as \emph{path} or \emph{route}.} called a \emph{mix cascade}.
Chaum proposes a deterministic node selection without specifying how the nodes are selected (node selection strategy) for mix cascades. 
He only suggests that certain factors such as the networks topology and user's trust can be used for mix node selection.
In a mix cascade, messages are successively encrypted (in a layered fashion) with the public key of each mix in the cascade (see Figure \ref{fig:mix}).

As the message is transferred from one mix to the next, the current mix peels off (decrypts) the corresponding layer (\ie remove one layer of encryption with its private key), obtains the inner layer together with the corresponding address of the next destination, and sends the message to that destination.
This procedure is repeated until the last mix delivers the data to its final destination.
In order to receive replies for messages while staying untraceable (to obtain recipient anonymity~\cite{Pfitzmann:2000:terminology}), return addresses are used. 
Chaum proposed to encrypt the address of the recipient of replies separately so that the respondent only needs to append the untraceable return address to her replies. 
The anonymous replies are also sent similarly in a layered fashion to the respondent.
From now on, we refer to the encrypted return address block as the reply block.
Note that in the case of the anonymous replies, the recipient of the reply is the routing decision maker. 

In order to overcome a single point of failure in availability of mix cascades, \emph{free-route} mix networks have been proposed.
In free-route mix networks, the route is not fixed and any sequence of nodes from the network can be used for relaying messages.
An important aspect in mix cascades and free-route mix networks design is how mixes are selected.
Selecting mixes for a mix cascade or for a path in a free-route mix network may follow different strategies. 
Namely, a deterministic strategy, a uniformly random selection, or a variation such as random selection biased by network state, or reputation/reliability scores.
When multiple mix cascades are available for the users to choose from, node selection has two dimensions:
selecting a set of mixes for building the cascades, and selecting a particular mix cascade for relaying the messages.
Moreover, predefined probability distributions and topological restrictions can also be taken into account for mix selection.
Danezis~\cite{Danezis:2003} proposed the \emph{restricted routes} mix networks that leverage the mix cascade model (\ie being less vulnerable to intersection attacks and being secure against global adversaries) and free-route mix networks (\ie being scalable).
He proposes a mix network topology that is based on constant degree graphs (sparse expander graphs), where each mix only communicates with a few neighboring nodes based on a predefined probability distribution.
Next, we review two variants of mix selection, one for free-route mix networks and one for mix cascades.

Mixes that fail, lead to further delays in mix networks, thus selecting reliable mix nodes can lead to better performance.
Dingledine \etal~\cite{Dingledine:2001:Reliability} proposed to identify mixes that fail and use a reputation system for mix selection leading to more reliability and efficiency for the mix network.
In their proposed system, mixes issue receipts for each received message. 
After a mix has sent a message to the next mix, if it is not receiving a receipt within a restricted time, it asks a set of witnesses to resend the message and receive the receipt and forward it to the original mix. 
The system establishes routing paths following the free-route node selection strategy, where the mixes are selected based on their past behavior (reputation score). 
Such a strategy suggests use of a non-deterministic node selection, biased towards mix nodes with high reputation scores.
Mixes that have no positive ratings at all are avoided for mix selection.
The main weakness of their scheme is that the reliability depends on the witnesses that need to be trusted, or at least a core group of trusted witnesses.

Unlike the previous system, which relies upon trusted global witnesses, Dingledine and Syverson~\cite{Dingledine:2003:Reliable-Mix} proposed a mix cascade protocol with distributed trust. 
The system they propose uses a reputation mechanism for rearranging mix cascades in order to obtain more reliable cascades. 
The construction of such cascade utilizes communal randomness and reputation scores provided by all of the mixes; therefore, there is no need of a trusted central authority.
To mitigate the weakness of the previous work, mix nodes of a cascade act as witnesses for the reliability of their own cascade.
All mixes submit random values to the configuration servers, which order mixes based on their reputation score and pick the top mix nodes to create a pool of mixes.
From this pool, the mixes are selected randomly of mix cascade rearrangement.
For each cascade, routing relevant information such as available bandwidth and expected waiting time are published. 
Based on this information and the reputation score of the mixes, users choose mix cascade for their messages. 
Note that if the mix network is large, the network view might not be complete for the users.


\subsection{Variations of Flushing Strategies}
Flushing algorithm (or batching strategies) specifies the precise timing at when a batch of collected messages is flushed out of the mix in order to be simultaneously delivered to the respective recipients.
Flushing strategies are analogous to the forwarding component of the routing and they highly influence the scheduling routing characteristic defined in Section~\ref{subsec:routing}.
Recall that Chaumian mixes collect messages until a certain threshold is reached such mixes are called threshold mixes. 
Threshold mixes might induce very high latency if the traffic load is low.
Thereafter, other flushing algorithms have been proposed in the literature.

Mixes that delay messages individually, for example based on a certain probability distribution, and lead to continuous flushing are called continuous mixes. 
One example of continuous mixes is the \emph{Stop-and-Go} mixes (\emph{SG-mix})~\cite{Kesdogan:1998:stopandgo} system. 
The initiator of a message assigns for each mix in the path 
a randomly selected delay (from an exponential distribution). 
The independent random delays that are assigned to each message make the performance and anonymity of each message independent of the other users in the system. 
However, a drawback of their system is that SG-mixes are vulnerable when incoming traffic is low \cite{Diaz:2004}.
Another type of flushing algorithms is pool mixes that only flush out a fraction of messages of a batch at each round, and keep the remainder in the memory of the mix (pool) for next flushing rounds. 
In pool mixes, the number of messages that are forwarded may be determined by deterministic or non-deterministic functions, and the message selection may be a uniformly random or weighted based on dynamic conditions (\eg based on incoming traffic). 
When the average delay of the messages is equal, pool mixes offer better anonymity since the anonymity set is bigger. 
Another advantage of pool mixes is that they are suitable for networks with fluctuating traffic load.
Pool mixes, however, still need to specify \emph{when} messages are flushed out and therefore combined with other flushing techniques such as threshold (described above) or time restrictions. 
Timed mixes enforce a time restriction for flushing out messages.
The anonymity of timed mixes is vulnerable to low traffic since if only one message arrives before the time restriction is met, the mix provides no anonymity measure for that message.
Moreover, a combination of the aforementioned flushing strategies can also be used by mixes~\cite{Diaz:2004,Serjantov:2004:Anonymity}.
For example, the two prominent \emph{remailers}, namely Mixmaster~\cite{moller:2003:mixmaster} and Mixminion~\cite{Danezis:2003:mixminion}, use \emph{timed dynamic pool mixes} as flushing strategies~\cite{Serjantov:2003:Attacks}, which are a combination of timed and threshold pool flushing techniques, where the parameters depend on the network traffic.
The flushing algorithm of Mixmaster has been characterized by generalized mixes \cite{Diaz:2003:Generalising-Mixes}. 
We review these remailer protocols in~Section~\ref{sec:remailers}.

Next, we review some mix protocols from the literature that have been suggested for applications such as ISDN telephone, web browsing, and anonymous emails.
In order to anonymize ISDN telephone communication with its intrinsic requirements on low-latency, Pfitzmann \etal~\cite{Pfitzmann:1991:ISDN-MIXes} introduced the concept of \emph{ISDN mixes}.
An important feature of ISDN mixes is to maintain constant traffic in the network to avoid traffic analysis. 
ISDN mixes use threshold mixes.
To obtain sender and receiver anonymity, ISDN mixes use two mix cascades, each built by the sender and receiver, respectively, which are connected either by a connecting mix; 
when used in long distance communications by the long distance network operators.
Initially, a broadcast takes place to exchange the connecting details and the time where the communication takes place.
To achieve constant traffic, a number of ISDN channels, with an equal amount of messages, need to start and end their communication at the same time (in a so-called \emph{time-slice}). 
However, this is time-consuming and would lead to blocking the connection, which is not suitable since ISDN mixes use narrow-banded channels and were designed for low-latency communication. 
In Table~\ref{table:mixdcnets}, we disregard the setup broadcast for exchanging connected information for ISDN mixes. 
The basic design of ISDN mixes was later generalized by Jerichow \etal~\cite{Jerichow:1998:realtimemix} to a system that enables low-latency, real-time communication.

A real-world realization built on ISDN mixes are \emph{Webmixes} (also known as JAP)~\cite{Berthold:2000:JAP,Berthold:2000:Anonymity} designed for real-time Internet applications, passing the traffic to several available mix cascades.
In Webmixes, the mixes transform the messages cryptographically and re-shuffle their order before flushing them out. 
However, messages are not delayed by flushing strategies. 
Webmixes use an adaptation of the time-slice method introduced by ISDN mixes. 
Routes in Webmixes consist of \emph{JAP proxies}, which are local software at the users, one (or several) mix cascade(s) consisting of reliable and high capacity mix nodes, and a cache-server.
Web requests are sent from the users JAP proxy through the mix cascade and the cache-server, and furthermore delivered to the destination server.
The web replies are sent back the same route and a copy of the reply is saved at the \emph{cache-server}. 
Hourly mix cascade information is published by so-called \emph{Info Servers}.
Users can choose among the published mix cascades by the info servers.
ISDN mixes, real-time mixes, and Webmixes have a deterministic node selection to build the mix cascade, where nodes selection for the cascades relies on the network state.

\subsection{Prominent Applications of Mixes: Remailers}
\label{sec:remailers}
The original concept of mixes has an immediate application to \emph{high-latency} remailer systems for providing anonymous e-mail service.

\emph{Babel} \cite{Babel:1996} aims at mitigating traffic analysis attacks by delaying only some messages of the batches. 
Babel uses independent forward routes and return routes. 
Forward routes may include a reply block (where the return route mix addresses are encrypted in a layered fashion) that may be used by recipients for anonymous replies.
Forward routes are considered to have better anonymity; one of the reasons for this is that reply blocks enable replay attacks on anonymous replies \cite{Danezis:2009:Survey}.
Babel introduces \emph{intermix detours}, where mix nodes choose a random sequence of mixes and relay the message through them before forwarding the message further to the next mix of the original route.
In Babel, the flushing algorithm uses time restrictions (intervals) and thresholds for flushing out messages.
Another technique Babel proposes to use is \emph{probabilistic deferment}, where a number of messages (determined by a biased coin) are delayed at each mix (this is similar to pool mixes). 
Babel proposes to use of free-route mix networks, where mixes are chosen uniformly random for each route by the user. 
However, there were no details given how routing information is communicated to users.
\emph{Mixmaster}~\cite{moller:2003:mixmaster} is an anonymous remailer, where mixes transform messages cryptographically into uniform sizes by adding random data at the end of each data packet. 
If a message is too large, Mixmaster splits up the message to achieve uniform sized packets and sends these packets independently of each other through a series of mixes, which do not necessarily need to be all the same.
Only the last mix needs to be the same for all packets of one email message, which has been split up before. 
Mixmaster adopts a free-route path selection, the node selection is not specified by the protocol, though statistics on the reliability of mixes can be used to bias node selection \cite{Danezis:2003}.
Though the Mixmaster protocol did not specify details about maintaining mix information, later implementations of Mixmaster adopted an ad hoc scheme for distributing routing information \cite{Danezis:2003:mixminion}.
One the main weaknesses of Mixmaster is that it only guarantees sender anonymity, since reply blocks are not used in Mixmaster.

\emph{Mixminion} (or Type III remailer)~\cite{Danezis:2003:mixminion} are widely considered as the state-of-the-art remailer. 
To guarantee equal routing information for all senders, Mixminion deploys a group of redundant and a synchronized system of \emph{directory servers}, which was not considered in the Mixmaster design.
Note that we disregard the directory servers synchronization for our classification in Table~\ref{table:mixdcnets}.
Like Mixmaster, Mixminion also uses ``timed dynamic pool''. 
Mixminion uses reply blocks. 
Generally, reply blocks enable replay attacks; hence, Mixminion introduces \emph{Single Use} Reply Blocks (SURB), where for each reply message, the content of the reply is appended to the SURB and sent through the mix network. 
In the Mixminion communication model, the routing path is divided into two so-called \emph{legs}, each consisting of half of the mixes in the route.
For reply messages, where both sender and receiver anonymity is desirable, in the first leg of the route, the sender of the reply chooses the mixes and appends the SURB for the second leg.
When the message is traversing the route, at a crossover point (the last mix in the first leg), the SURB replaces the first leg, and the message is routed further to the intended recipient.
In such cases, the route consists of mixes, which are half chosen by the sender and the other half chosen by the recipient. 
Thus, Mixminion aims at providing sender anonymity and recipient anonymity for email messages.
Moreover, since forward and reply messages are not distinguishable from each other by outsiders and intermediate mix nodes themselves, they share the same anonymity set.
The exceptions are the crossover points that have partial knowledge and the exit mix nodes because they can observe whether the content has been encrypted or is in plain text.
Mixminion also suggests choosing nodes from preferably a large pool; however, further details on the node selection strategy have not been specified in Mixminion. 



\subsection{Onion Routing}\label{sec:onionrouting}

Onion routing \cite{Goldschlag:1996:onionrouting}\cite{Reed:1998:Onion-Routing} is designed for anonymizing connections for applications with low-latency constraints, such as web browsing.  

\begin{figure}[htp!]
\centering
\includegraphics[width=.36\textwidth]{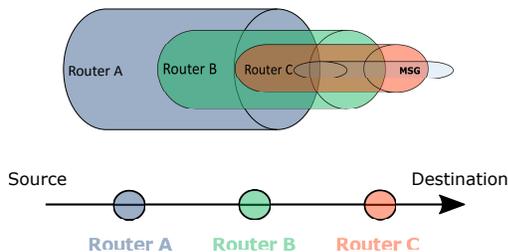}
\caption{The concept of onion routing}
\label{fig:onion}
\end{figure}

An onion routing network consists of a set of nodes so-called \emph{Onion Routers (ORs)}.
Users choose an ordered sequence of ORs to establish a bidirectional channel, so-called \textit{circuit}, for relaying their data through the onion routing network.
The communication is encrypted in a layered fashion and the ORs in the circuit each can decrypt their corresponding layer. 
When the communication is relayed by an OR in the circuit, the OR removes the corresponding layer of encryption and forwards the data to the next OR in the circuit (see Figure~\ref{fig:onion}). 
The last OR forwards the data to the destination. 
Each OR only knows their predecessor and successor in the circuit, and the complete sequence is only known to the circuit initiator (the user). 
Therefore, only the first OR in a circuit is aware of the IP address of the user who has initiated the circuit and only the last OR of a circuit is aware of the destination of the communication, which is relayed through the circuit. 
The response of the receiver is relayed back to the initiator through the same circuit. 
Similar to Webmixes, in onion routing, the ORs implement First-In First-Out (FIFO)-like forwarding strategy to provide low-latency services. 
Having no delays at the ORs and due to missing cover traffic onion routing are susceptible to a number of attacks, such as traffic analysis and timing attacks, where the adversary may identify and correlate traffic patterns at the initiator and receiver~\cite{Danezis:2008:Survey, Ren:2010:Survey}, thus de-anonymizing the connection. 
Nonetheless, onion routing is a promising design to provide a low-latency AC network, and many currently used systems to build upon this design.

\subsection{Onion Routing-based Protocols}
\label{sec:onionroutingbased}
Onion routing is used in Tor \cite{Dingledine:2004:Tor}, which constitutes an extension of the original onion routing design, with some modifications to achieve better security, efficiency and deployability. 
The Tor network, an open source and free to use the framework, consists of a large set of volunteering routers (at the time of writing, there exist more than 7000 routers~\cite{tormetrics}).
The network is mostly connected because routers can connect to any router from the Tor network, except for connections between routers located in the same IP /16 subnet space, which are not possible. 
Tor's services are used daily by approximately 2,000,000 users~\cite{tormetrics}. 
Each user runs a piece of software called Onion Proxy (OP) that manages all Tor related processes, \eg establishing circuits or handling connections from user applications. Tor deploys a group of well-known and trusted authoritative servers that publish on a regular basis (typically, every hour) a list of all active Tor nodes with their characteristics, \eg estimated bandwidth, IP addresses, and cryptographic keys. 
This list is called a \emph{consensus}.
After the user has obtained the consensus, the OP of the user chooses an ordered set of usually three ORs to build a circuit.
The first node in a circuit is called the \emph{entry} node, the second node is the \emph{middle} node, and the last node in the circuit is the \emph{exit} node. 
The first node that is selected is the exit node, then the entry node of the circuit is selected, and last the middle node of the circuit is selected.
After selecting a set of ORs, the OP contacts the entry node and builds a circuit with it. 
This newly created circuit is used to contact the middle OR to extend the circuit and similarly through the middle node the exit node is contacted to extend the circuit. 
The established circuit can now be used to anonymously relay data. 

In 2002, Wright \etal introduced the predecessor attack \cite{Wright:2002:Degradation} on onion routing. 
To defend against this and related attacks, selecting a small set of nodes was introduced for Tor \cite{Wright:2003:defending-predecessor-attacks}.
Previously, each user maintained a list of 3 randomly pre-selected (so-called \emph{guard}) nodes with high bandwidth and uptime. 
This list was updated every 30/60 days and the user could choose uniformly random an entry node from this list for each path construction.
This has changed recently because Tor is starting to let each user select only one fixed entry guard node for 9 months \cite{Dingledine:2014:OneGuardTor}. 

In the early onion routing design, it was suggested to select the nodes uniformly random \cite{Syverson:2001:OR-Security}. 
Due to performance considerations, Tor's routing policy does not select nodes with the same probability, but rather preference is given to high-bandwidth nodes. 
The likelihood that nodes are chosen for certain positions in a given route depends on the ratios of overall node bandwidths and node such as the IP addresses and whether they can be selected as entry node or as exit node. 
Moreover, some additional bandwidth weights are used to balance off the node selection. 
As mentioned before, a further development in the routing policy is to disallow a communication to pass through two nodes within the same /16 subnet IP address.
The implications of these changes with respect to structural node corruption have been recently explored by Backes \etal \cite{Backes:2013:AnoA,Backes:2014:Mator}. 

Next, we review two prominent attacks on Tor's routing.
Murdoch \etal have proposed a traffic-analysis attack using timing information to identify Tor nodes and to infer traffic load to a specific initiator. 
Their investigation shows a degradation of Tor's anonymity against such attacks. 
They furthermore propose some strategies to prevent the risk of such attacks, mainly by increasing communication latency~\cite{Murdoch:2005:Traffic-Analysis}. 
Bauer \etal have proposed a traffic analysis attack aim at decreasing the anonymity of Tor \cite{Bauer:2007:Attack}. 
Their attack investigates the load balancing that is performed by Tor, where high bandwidth nodes are preferred in the node selection strategy. 
They show that performance optimization impairs the anonymity of Tor against end-to-end traffic analysis attacks.

Since Tor has been proposed, there has been a great deal of research on extending Tor's routing strategy. 
The proposed extensions to the Tor routing protocol aim mostly at improving either the achieved anonymity of Tor, or the performance that Tor users experience. 

Improvements to Tor's anonymity have been often realized by aiming at an improved node selection. 
For example, improving anonymity by using  better weighting at the node selection phase has been proposed in~\cite{Panchenko:2012:ImprovingTorPerformance} and \cite{Backes:2014:Mator}. 
Involving AS-level information in the node selection has been proposed by \cite{Edman:2009:Tor-AS-Aware} and \cite{Akhoondi:2012:LASTor}. 
Moreover, offering the user a tuneup option between uniformly random node selection (for high anonymity) and weighted random node selection with a bias towards high bandwidth nodes (for better performance) has been suggested by Snader and Borisov \cite{Snader:2011:Tor-tuneup}.

Tor's performance problems have several causes, and hence suggested improvements aim at different aspects of the Tor routing protocol. 
One cause of Tor  performance is high congestion \cite{AlSabah:2015:Tor-Survey, Dingledine:2009:TorTechReport}, often caused by bulk traffic, which induces high latency for interactive/web traffic. 
Several solutions to solve the problem of high waiting times for interactive traffic have been proposed. 
One possible solution is to increase the number of connections between two nodes \cite{Geddes:2014:Imux, AlSabah:2013:PCTCP, Gopal:2012:Torchestra, AlSabah:2012:Conflux}, which can be used to separate interactive and bulk traffic into different connections.
Another solution is to prioritize interactive traffic in the scheduling phase \cite{Tang:2010:Tor-CircuitScheduling} \cite{AlSabah:2012:DiffTor}. 
An alternative solution is to improve how Tor's resources are used 
by improving node selection with a more realistic estimation of the available bandwidth of nodes \cite{Panchenko:2012:ImprovingTorPerformance}. 
Furthermore, another solution to Tor's congestion problem is to enforce avoiding congested nodes at the node selection phase \cite{Wang:2012:Tor-CongestionAware}. 
Another reason for Tor's high latency is circuitous paths \cite{Akhoondi:2012:LASTor}. 
To solve this problem, node selection strategies have been proposed that take the destination between chosen nodes into account \cite{Akhoondi:2012:LASTor, Sherr:2009:Tor-Link-BasedSelection, Panchenko:2012:ImprovingTorPerformance}.  

The scalability of Tor has also been subject to new proposals for the Tor routing protocol in the literature. 
One proposal to tackle scalability issues is to give the user only the information about the necessary nodes for path construction and to hide the complete view of the system from the user by either managing Tor nodes as a DHT table and using Kademlia for node retrieval \cite{McLachlan:2009:Torsk}, or by using private node retrieval \cite{Mittal:2011:PIR-Tor}.

\subsection{Random Walks, Structured and Unstructured DHT-based Protocols}\label{sec:dht}

In this section, we review \emph{random walk protocols}, where the communication is relayed randomly through the network. 
We consider a protocol a random walk protocol if node selection is hop-by-hop routed and a random selection.
Random walk protocols are often combined with peer-to-peer network structures.

\label{sec:crowds}

\emph{Crowds}~\cite{Reiter:1998:Crowds} is one of the early AC systems designed for anonymous web browsing. 
The key design feature of Crowds is a random peer selection.
In Crowds, all nodes are grouped into so-called \emph{crowds}; all nodes within a crowd might connect to each other for relaying a communication. 
Each node in the crowd is called a \emph{jondo}. 
A so-called \emph{blender} is responsible for managing and administrating nodes. 
Crowds has a peer-to-peer structure since all users of the system are nodes themselves. 
The user randomly selects a node and sends her message (\ie website request). 
Upon receiving the request, this node flips a biased coin to decide whether to send the request directly to the receiver or to forward it to another node selected uniform at random. 
This continues until the message arrives at the destination. 
The server replies are relayed through the same nodes in reverse order. 
Wright~\etal showed that Crowds is vulnerable to so-called predecessor attacks \cite{Wright:2002:Degradation, Wright:2004:PredecessorAttack}. 
In order to prevent such type of attacks, Crowds suggested to employ static route (a user keeps the route for a while) such that an attacker does not have multiple routes to link to the same jondo~\cite{Reiter:1998:Crowds}. 
However, even keeping routes static for a day is not enough to prevent predecessor attacks~\cite{Danezis:2009:Survey}.

\emph{MorphMix}~\cite{Rennhard:2002:MorphMix, Rennhard:2004:MorphMix} is a dynamic peer-to-peer AC network. 
Technically, MorphMix establishes circuit-based connections using layered encryption, where the anonymous route is established iteratively by the nodes on the route.
Each node is typically only aware of a set of network nodes, which is not necessarily covering all nodes.
In order to avoid repeated connections with the same set of nodes, a node has to forget about nodes it has not been connected and constantly require new node information.
After an initiator selects the first node, she selects randomly a witness for each hop thereafter, randomly chosen from the nodes in her local database.
She asks the next hop to extend the route with the assistance of the witness she has chosen, where nodes propose a set of candidate nodes for the next hop and the witness chooses one of them.
To prevent path compromise, 
nodes can only propose nodes with different IP prefix to her own IP address to the witness. 
The witness should not be selected from the nodes to which the initiator is connected currently to avoid initiators being identified by witness nodes. 
In order to mitigate guessing whether a node was initiator by the next hop, the initiator adds random delays to her communication before forwarding them in the tunnel establishment phase.

Efficiency is one of the main problems in random walk protocols. 
In the next section, we review DHT-based protocols, which aim at efficient node lookup and selection. 
Random walk protocols can employ DHT lookups to gain better efficiency (\eg AP3 protocol~\cite{Mislove:2004:AP3}). 

\subsubsection{DHT-based Protocols}
\label{sec:dhtrouting}
\begin{figure}[htp!]
\centering
\includegraphics[width=.36\textwidth]{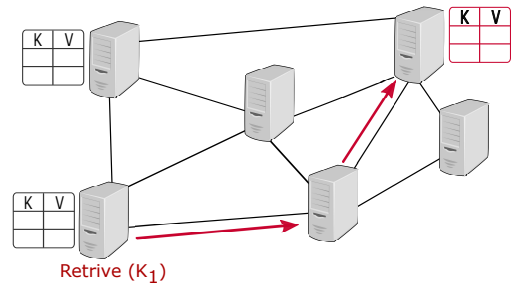}
\caption{The concept of distributed hash tables}
\label{fig:dht}
\end{figure}

In distributed systems, where there are network administrators, a challenge is to locate a node. 
One solution is to use Distributed Hash Tables (DHTs) to manage the distributed nature of the data (relaying nodes or distributed storage).
Generally, DHT refers to a trust-distributing, structured-data management model for storing (value, key) pairs and is accompanied with key-based lookups for locating the corresponding stored value (see Figure~\ref{fig:dht}). 
The value might be, for example, either the router information of relaying nodes in a distributed network or a stored content (file).
The keys are hashed from the identifier of the value (for nodes, their IP addresses are hashed into keys). 
In the literature, several lookup strategies for the DHT-based structures have been proposed, aiming at efficient searching. 
Some popular lookup strategies are Kademlia \cite{Maymounkov:2002:Kademlia} (locating the nodes based on their estimated distance using an XOR metric), Chord \cite{Stoica:2001:Chord} (using a clockwise circle metric, where at each hop of the lookup, the distance to the node is decreased, at least half), and Pastry \cite{Rowstron:2001:Pastry} (carrying out lookups based on numerical identifiers).

DHT structures enable efficient routing even when the peers of a DHT structure keep only information (key-value pairs) of a partial subset of all the other peers of the DHT structure; 
this, in turn, leads also to improved scalability of such systems. 
Another important feature of DHT-based structures is having better load balancing.
For systems, where nodes have only a partial view of the structure, hop-by-hop routing is preferable. 
Some AC protocols use randomness in the routing strategy besides the classical lookup method. 
For example, node selection is carried out by selecting a random key and by then using a classical lookup method (an adaptation of Chord, Kademlia, or Pastry) to find that key.
Next, we review AC protocols that use an adaptation from Kademlia, Chord, Pastry for their node lookup (considered as structured DHT-based protocols).
We proceed by reviewing independent DHT-based routing proposals for AC that are considered unstructured DHT-based protocols.
We start with \emph{AP3}~\cite{Mislove:2004:AP3}, a random walk protocol aiming at providing anonymity when a large part of the nodes is compromised. 
AP3 uses the same routing strategy as Crowds, with the difference that the node information is retrieved using Pastry and that the node does not have a complete view of the system.

Next, we review two protocols that aim at replacing node selection of source-routed protocols such as onion routing with structured DHT systems making the suitable to be combined with onion routing. 
\emph{Salsa}~\cite{Nambiar:2006:Salsa}, proposed by Nambiar~\etal, aims at providing scalability and preventing malicious colluding nodes to be able to bias routing. 
Salsa virtually divides nodes into groups, which are organized in a binary tree form. 
For routing, simultaneous redundant lookups and bound checking are used in order to avoid malicious nodes returning wrong addresses. 
The lookup queries are carried out similar to the Chord lookup in a recursive fashion. q
In Salsa, the routing information that is available to each node is partial; however, the tree structure allows nodes to carry out source-routing. 

McLachlan~\etal have proposed \emph{Torsk}~\cite{McLachlan:2009:Torsk}, a peer-to-peer AC protocol, replacing Tor's node selection and directory service with a DHT design. 
It aims at providing better scalability for Tor. 
Their design uses DHT tables for node selection by using a randomly chosen key that is looked up in the table using Kademlia. 
To secure lookups, Torsk uses the ``root certification'' approach proposed by Myrmic \cite{Wang:2006:Myrmic} and randomly selected secret ``secret buddies.''

Panchenko~\etal proposed \emph{NISAN}~\cite{Panchenko:2009:NISAN}, an AC protocol that aims at achieving high scalability and preventing adversaries to bias routing. 
NISAN uses iterative search to select nodes randomly for constructing anonymous paths.
It uses an adaptation of Chord, where the node lookups are aggregated. 
Moreover, NISAN hides the node that it is looking up, by requiring the complete routing table and enforcing bound checking to further prevent selecting nodes from routing tables, which were manipulated by malicious nodes. 

\emph{Octopus}~\cite{Wang:2012:Octopus} aims at providing security by preventing malicious nodes to be able to bias routing. 
It also aims at providing anonymity by hiding which nodes have been looked up for anonymous paths. 
For routing, Octopus uses iterative lookups by sending the query to the closest node to the searched key in the local routing table and then retrieving the routing table from that node until the node containing the corresponding value to the key is found. 
Node selection is carried out in two phases. 
In the first phase, nodes are selected by the path initiator (user).
In the second phase, the last node selected in the first phase chooses the remaining nodes. 
Therefore, Octopus is not purely a random walk protocol.
After establishing anonymous paths, Octopus splits queries to different paths and adds dummy traffic to hide the real queries among them. 
Furthermore, as security measures, Octopus enforces bound checking on the received routing tables to prevent using manipulated routing tables, and it proactively tries to identify and remove malicious nodes.

Next, we review two file sharing protocols that use DHT for routing file requests and their responses. 
They, however, use unstructured routing. 
Clarke~\etal proposed \emph{Freenet}~\cite{Clarke:2001:FN}, a peer-to-peer censorship-resistant system for sharing storage space. 
Freenet offers strong decentralization in order to provide privacy and robustness against attacks. 
The key design feature of Freenet is based on storage replication and plausible deniability. 
Files are stored multiple times at the nodes, are indexed by binary file keys, and can be looked up by their corresponding key. 
Each node has a dynamic routing table including the node information with the stored keys. 
The original design uses a heuristic deterministic routing using potentially all Freenet participating nodes choosing mostly neighborhood nodes (currently called Opennet mode). 
Freenet uses an adaptive routing using DHTs with keys that are location-independent. 
Three methods are used for key construction: keyword-signed key, signed-subspace key, and content-hash key (for more details see~\cite{Clarke:2001:FN}). 
The routing table is updated periodically to achieve better performance. 
The replication of files provides resilience against node failure and node overloads.
In the Opennet mode, a heuristic-based deterministic routing approach (a distance-directed depth-first search with backtracking) is used \cite{Clarke:2010:FN, Roos:2014:FN}. 
When a file request arrives at a node, including a key and a value for hops-to-live, if the file is not stored locally, the node looks up the node with the nearest key in the routing table and forwards the file request to the corresponding node. 
The node receiving the request repeats the process until either the file is found or the hops-to-live is reached. 
If the requested file is found, the node forwards the file to the node from which it has received the request, stores a copy of the file locally and updates its routing table in order to optimize routing for future requests. 
If the node that is contacted is not responding, the node sends the request to the node with the second-nearest key. 
If that node is also unresponsive, it contacts the third-nearest one, and so on. 
If the file is not retrieved within the hops-to-live number of hops then the search is aborted and the file requester is informed. 
The nodes that are forwarding the requested file back to the file requester change randomly the sender address, providing reasonable deniability for the node that has stored the file~\cite{Clarke:2001:FN}.
The Opennet mode was vulnerable to various attacks. 
In particular, nodes participating in Freenet were not protected, and an attacker could easily find out whether a router is a participating Freenet node. 
In the Darknet mode, such shortcomings are addressed. 
In 2010, Freenet has been extended by a membership-concealing Darknet mode, where trusted connection are used for routing \cite{Clarke:2010:FN}. 
In the Darknet mode, the user chooses the nodes from her trusted nodes \cite{Clarke:2010:FN}. 
The routing table is consisting of nodes derived from a fixed graph, which is the social graph of the node. 
In the Darknet mode, the routing table is not optimized during time and cannot include nodes that are not derived from the social graph of the node. 
Since the Darknet mode is based on the trusted network of a user, the structure of the network is following Kleinberg's small world model \cite{Kleinberg:2000:SWM}.
The relaying nodes only know their predecessor and the successor in order provide privacy. 
In Freenet, the data is encrypted using symmetric encryption. 
The key is transferred either with the address or in the header of the file request~\cite{Clarke:2001:FN}.

\emph{GNUnet} \cite{Bennett:2002:GN} was originally designed as a peer-to-peer censorship-resistant content sharing system, 
but has been expanded into other applications such as anonymous file sharing using the \emph{GAP} protocol~\cite{Bennett:2003:GAP}. 
GAP aims at providing requester and responder anonymity for file search and file sharing.
In GAP, a node that is relaying a message in the forward route has the option to ``drop out'' from the reply route (for example due to network state and its own heavy load) and when the reply is sent back, the node is over-jumped. 
Moreover, when queries arrive at the nodes, they can be dropped if the node has already too much load.
Routing in GAP uses credit rating scheme, where relaying requests and replies increases the credit and sending uses the credit. 
The credit score is local at each node.
In GAP, the file request can either be sent to newly selected nodes or to a node where there is already a connection established. 
This is decided based on the node's current CPU and load, the credit rating and a random factor.
The node selection is random with a bias towards nodes, which have a closer identifier to the hash value of the file that is queried.
Moreover, the network activity is also taken into account in node selection (giving preference to ``hot paths''). 
Unlike Freenet, GAP uses a time-live restriction to avoid routing loops; when time-to-live is reached, the query is forwarded directly to the destination with a certain probability.
For flushing in GAP, nodes use a combination of timed and threshold mixes for flushing batches of messages, where the time restriction is selected randomly.

\subsection{DC Networks}

\label{sec:DCNets}

\begin{figure}[htp!]
\centering
\includegraphics[width=.36\textwidth]{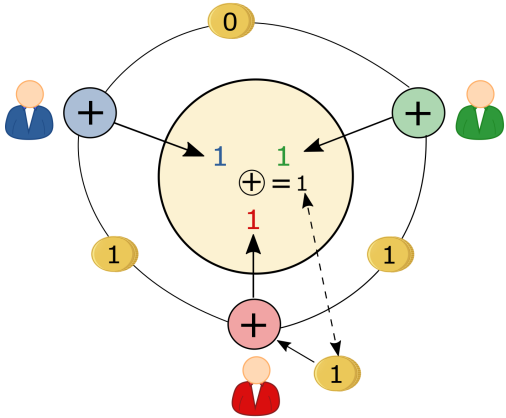}
\caption{The concept of DC network}
\label{fig:dcnets}
\end{figure}

The idea of \emph{DCnets}~(Dining Cryptographers Networks) was first proposed by Chaum~\cite{Chaum:1988:DCNets} and later revisited~\cite{Golle:2004:DC, Waidner:1990:DC}. 
DCnets are an important alternative to mix-based schemes and their extensions due to their resistance against traffic analysis attacks. 
DCnets offer non-interactive anonymous communication using secure multi-party computation with information-theoretically secure anonymity, guaranteeing sender anonymity while enabling all participants to verify the final outcome.
The key concept is that every participant outputs a message that is disguised by XORing them with the keys the participants are sharing pairwise with other participants.
The participants combine their outputs and share the output with each other (\ie they broadcast their output).
When the encrypted messages are combined, the keys cancel each other out, and the message is revealed; however, the sender remains unknown (see Figure~\ref{fig:dcnets}).

The DCnet concept can be generalized, to transmit large messages simply by repeating the protocol as desired~\cite{Golle:2004:DC}.
DCnet expects all participants to be involved in every run of the protocol and requires pairwise shared keys between the participants.
Moreover, every participant needs to disclose the same number of bits in each round.
The participants can share the keys for every round, or they can repeatedly use the same key;
this makes DCnet unconditionally or computationally secure, under the assumption that the protocol is executed correctly.
Moreover, DCnets also have practical challenges, such as the message transmission or avoiding collisions (unintentional) and disruptions (intentional collisions).
Since a collision invalidates the message (bit), when only one-bit messages are sent, just one of the participants may transmit at a time (although all participants are involved in each round). 
If multiple participants want to send messages within a block of communication, they need to occupy different positions within the block.
One proposed solution is to randomly pick a position (slot) in the block that is going to transmit and reserve the position in earlier rounds (pre-transmission round).
However, this might only shift problem and again in the reservation round collisions might occur.
The basic DCnet does not prevent any disruption, such as actively blocking participants from sending the message; hence, it is susceptible to anonymous DoS attacks.
To partially address this problem, some solutions to detect disrupters in DCnets have been proposed in the literature \cite{Bos:1990:DC-Disrupters, Waidner:1990:Attack}.
Furthermore, recovering from a fault is only possible by re-broadcasting the messages.

Chaum proposed in his DCnet to either use a ring topology for sharing the messages or use broadcast to transmit messages to all participants at once.
The ring topology solution has a the problem of detecting the disruptions because malicious participants can adapt their answers to other participants answers to avoid being detected. 
Basically, if two users submit reverse bits, they cancel each other out and the disruptions remain undetected.
Other topologies that have been proposed for DCnets are tree \cite{Dolev:2000:DCnet-XORTrees} or star topologies \cite{Pfitzmann:1986:Unobservability}.
The broadcast solution has the problem of being expensive and introduces the problem of collision. 
The major limitations of DCnet are the strong assumptions that they require:
first, participants follow the protocol honestly and are expected not to collude;
second, unconditional sender anonymity is guaranteed only if there is an unconditional secure channel between every pair of participants.
Furthermore, DCnets are vulnerable to Sybil attacks \cite{Douceur:2002:Sybil}.

\emph{Herbivore}~\cite{Goel:2003:Herbivore} is built on top of DCnets aiming at better efficiency and scalability and managing churn.
To improve scalability, Herbivore breaks down the participants into smaller groups called \emph{cliques}, a message can only be traced to a clique but not to the corresponding sender/receiver within their clique.
Within a clique, participants are organized in a star topology, where the central node relays all messages between members of a clique. 
The central node is changed for each new round of communication.
For inter-clique communication, the cliques are connected to each other in a ring topology.
For locating cliques, Herbivore employs the Chord protocol \cite{Stoica:2001:Chord}.
In order to mitigate intersection attacks, nodes departure from a clique can be vetoed by the node that is in the middle of a long-run transmission.
Although authors claim low-latency, we decided to classify the protocols as being \emph{high latency} since it contains a central node that has to wait for messages from all other nodes in the clique. 
One of the main weaknesses of Herbivore is that smaller anonymity sets are achieved and the applications have a time restriction based on the cliques lifetime.
Moreover, the star topology makes the design vulnerable to DoS attacks.

Dissent (Dinning-cryptographers Shuffled-Send Network)\cite{Corrigan-Gibbs:2010:Dissent} is a latency-tolerant protocol for AC. 
It is the first protocol that provides accountability for a small-size group, and also maintains integrity. 
Dissent is built on top of DCnets, but relaxes the aforementioned assumption that all participants follow the protocol correctly.
In Dissent, anonymous communication is guaranteed for members of a group.
Apart from the multi-party computation and layered encryption to hide the sender of the messages, to solve the collision problem, each group member influences the position of the messages of other group members in the final transmission block. 
Dissent consists of two sub-protocols: a \emph{shuffle protocol} and a \emph{bulk protocol}, 
In the bulk protocol, each member creates an assignment table for each of the other member, so-call \emph{message descriptors}. 
The shuffle protocol is used to shuffle these messages descriptors.
Based on these message descriptors, each participant inserts her messages to a cipher stream, which is a slice of the message block that needs to be transmitted. 
The shuffle protocol functions similar to mix cascades, where each participant receives the set of message descriptors (which were encrypted in a layered fashion) and shuffles them and passes them over to the next participant.
Thereafter, each member transmits one cipher stream.
When these cipher-streams are combined, a vector of concatenated messages is obtained. 
Dissent uses broadcasting for intermediate runs of its protocols such as sharing keys. 
However, the final cipher streams are not necessarily broadcasted, and can be sent to a single group member or a non-member node. 
Hence, Dissent primarily only guarantees sender anonymity and further protocol setup details determine whether recipient anonymity is also achieved.
To mitigate untraceable DoS attacks (disruptions), go/no-go messages and blame phases are used in Dissent, which identify collisions and malicious participants and enables accountability.


Wolinsky \etal have extended Dissent to improve scalability and efficiency \cite{Wolinsky:2012:Dissent}.
They propose to group participants and use designated servers, where the group members share keys with these servers instead of each other (the network consists of server nodes and participant nodes).
In the basic version of Dissent, the group size was restricted; however, in the extended version, the participants may form larger groups, though the servers consist of a significantly smaller group, while still being not completely centralized to avoid the single point of failure.
Hence, the extended Dissent builds an asymmetric topology for key sharing.
At least one of the servers needs to be honest to prevent compromises. 
While latency introduced at the shuffle protocol made the basic version of Dissent unsuitable for interactive and low-latency applications, the extended Dissent, if used in a local-area setting, can be suitable for low-latency communication.







\subsection{Miscellaneous Protocols} 
\label{sec:misc}

\emph{Tarzan}~\cite{IPTPS:2002:Tarzan,Freedman:2002:Tarzan} is a peer-to-peer anonymous fully decentralized IP-level network overlay. 
All participants are peers; they are all potential originators of traffic, and also potential relays. 
Tarzan nodes do not implement any mixing strategies and simply forward incoming traffic.
After the initiator node selects a set of nodes (preferably from existing connections from previous communication rounds) to form a route through the overlay network, a tunnel via these nodes is established for relaying communication. 
Unlike the early design of the protocol~\cite{IPTPS:2002:Tarzan}, where the peers only needed to know about a random subset of nodes,
the final version~\cite{Freedman:2002:Tarzan} introduces a \emph{gossip-based} protocol based on the Name-Dropper protocol~\cite{NameDropper:1999}, where more node information is requested from randomly chosen nodes. 
The aim is to gain information about all available servers in the network to avoid attacks that are facilitated due to churn, such as fingerprinting attacks~\cite{Danezis:2006:Fingerprinting}.
Node information is stored in a ring model and lookups are carried out using the Chord algorithm~\cite{Stoica:2001:Chord}.
The initiator only selects nodes randomly from distinct IP subnets, a three layer hierarchy selection is used to make sure nodes are from distinct subnets. 



\emph{I2P}~\cite{Timpanaro:2012:I2P} is a distributed overlay network, originally aimed at enabling anonymous communication between two nodes within the I2P network.
Note that currently there is a service built on top of I2P to allow getting connected to web servers~\cite{Timpanaro:2011:MonitoringI2P}. 
Currently, the number of I2P routers is estimated to be between 40,000 and 50,000 \cite{I2P:2016:stats}.

The network metadata (containing router contact information and destination contact information) is distributed among a subset of all nodes so-called \emph{floodfill} nodes, and is managed using DHT structure by employing Kademlia for node lookups. 
At bootstrapping, users obtain a list of I2P peers from websites and then contact two floodfill routers from the list and requests router information that is available to that floodfill node. 
In order to mitigate that malicious floodfill nodes are not biasing node selection by providing manipulated router information, router information is stored at eight floodfill nodes~\cite{Egger:2013:I2P-Practical-attacks}.

Nodes are categorized into tiers (called \emph{peer profiling}) based on the previous performance (response times) and reliability (uptime) of nodes. 
Three main types of tiers are defined in I2P: high capacity, fast, and standard. 
The routing protocol of I2P, so-called \emph{Garlic Routing}, is source-routed with a randomized node selection biased towards faster nodes~\cite{Schimmer:2009:I2P}. 

In I2P, communication channels are unidirectional and called tunnels; tunnels for outgoing traffic are called \textit{outbound} and tunnels for incoming traffic are called \textit{inbound}. 
Each user maintains a number of inbound and outbound tunnels; outbound inbound tunnels of other users can be retrieved from the floodfill nodes.
When users want to relay communication to each other, the nodes in the chosen inbound and outbound tunnels shape the relaying route. 
Moreover, there are two types of tunnels in I2P -- client tunnels and exploratory tunnels -- for which different peer selection strategies are used. 
Client tunnels are used for application traffic, and exploratory tunnels are used to send administrative information. 
For client tunnels, peers are selected randomly from the nodes that are categorized as fast-tier nodes, which is done locally by the client using previous measurements. 
For exploratory tunnels, peers are selected randomly from the set of nodes that are categorized as standard tier. 

The communication through I2P is protected using \textit{garlic encryption}. 
Garlic encryption is very similar to onion encryption, with the difference that multiple data messages may be contained in a single \textit{garlic message}. 





\section{Discussion}
\label{sec:discussion}

\subsection{Routing Features: Commonalities and Differences}
In this section, we discuss commonalities and differences between the investigated classes of AC protocols with respect to their routing characteristics.
The discussion is grounded on our classification presented in Tables~\ref{table:mixdcnets} and \ref{table:onionrouting}, and strives to provide a deeper understanding of the relationships of individual routing characteristics.


\emph{Mixnet-based protocols}, as classified in Table \ref{table:mixdcnets}, show the most heterogeneous routing design among the four investigated protocol classes.
The main reason for this is they demonstrate routing diversity on multiple routing building blocks, such as proposing disparate flushing strategies, differentiating node selection strategies, which in turn lead to topological differences.
As mentioned earlier, existing routing strategies can be classified into \emph{free-routes} mix networks and \emph{mix cascades}. 
However, we distinguish whether a connection is potentially allowed between two nodes or not based on routing of the messages. 
Hence, we marked most of the mix cascade networks as fully connected and only Webmixes and Restricted routed mix networks as partially connected.

Generally speaking, mix cascade networks employ rather synchronized connection because messages are sent in batches and mostly depend on their flushing algorithms in a timely schedule. 
For example, timed mixes lead to synchronized message transmission.
Recall that the flushing algorithm in Mixmaster and Mixminion partially uses time restrictions.
However, we consider these two protocols with asynchronous message transmissions due to the possibility that low traffic might lead to a threshold restriction instead of a time restriction.
As for free-route systems, in SG-mixes, message transmission is also synchronized due to assigned time ranges by the routing initiator.
Nevertheless, these timing ranges are not coordinated with other users or mix nodes. 
Dingledine \etal's proposal for a reputation system for mixnets \cite{Dingledine:2001:Reliability} also uses a synchronized message relaying to enable verifying the correctness of the routing process.

In the mix protocols, node management has not been always specified in the protocol description. 
For example, in Chaumian mixes, the view of the routing decision maker is not discussed; however, it can be implicitly deduced that it is complete. 
The anonymous remailer Mixmaster does not discuss node management either; however, the later implementation uses ad hoc systems, which suggests a partial view~\cite{Danezis:2003:mixminion}. 
The remailer Mixminion defines a node management strategy to insure a complete view for the routing decision maker.

Source-routing is one of the inherent routing features of mix cascade protocols because the routing paths are fixed beforehand. 
Choosing the mixes for the mix cascade might be either deterministic such as in the case of Webmixes or non-deterministic such as in the case of Reliable mix cascades.

Flushing algorithms do apparently impact scheduling.
Note that some protocols in Tables~\ref{table:mixdcnets} and~\ref{table:onionrouting} use randomness in the scheduling process (\eg pool mixes).
Consequently, some messages are forwarded later than others. 
Since individual messages do not have priorities by themselves, we categorized them also as fair.
How the set of nodes is derived for node selection has also not been specified precisely for mix networks.
The same holds for selection probability, such as for Chaumian mixes.
For mix networks, we categorized the selection probability as deterministic because all mixes are chosen for a single mix cascade.
For both mix cascade protocols and free-route mix networks, the selection set varies depending on the application of the AC network and on the potential anonymity properties. 

As mentioned in Section \ref{sec:mixnets}, in mix cascades, the selection probability has two dimensions when more than one cascade exists. 
For instance, Webmixes can provide multiple mix cascades, where mixes are chosen by the network administrator for each mix cascade. 
Thereafter, the user manually selects one of these mix cascades for routing her messages. 
Another mix cascade protocol, where mixes are selected deterministic, is ISDN mixes.

All mix cascade protocols are high latency AC networks and have a message-based communication mode;
exceptions are ISDNs, Real-time mixes, and Webmixes, which are designed for low-latency applications, such as web browsing. 
Note that the latencies might be restricted, for instance in case of Stop-and-go mixes, where the delays are randomly selected from a restricted time range. 


\emph{Onion routing protocols}, as classified in Table \ref{table:mixdcnets}, are all Tor related schemes and hence, exhibit the most homogeneous routing design among the four investigated protocol classes.
On a conceptual level, all these protocols are equally characterized by their routing features.
However, there are three exceptions that affect: the completeness of the network view, the fairness of scheduling, and the node selection probability (leaving apart the non-technical question if the code has been made publicly available).
Their differences, however, often lie in implementation details, which are not necessarily relevant to routing, such as reducing buffer size \cite{Alsabah:2011:DefenestraTor}. 
Also, differences in the routing policy, which do not change the routing feature on a conceptual level such as changing node selection probabilities \cite{Backes:2014:Mator} and \cite{Panchenko:2012:ImprovingTorPerformance}, are equally classified in the table, though node selection probabilities could be different.

One inherent routing feature of onion routing protocols, due to preventing additional latency, is to have no synchronization, which makes such protocols sensitive to timing attacks and global adversaries.
Another inherent feature is that all onion routing protocols have a client-server model, which improves their usability and leads to a higher number of users, thus contributing to better anonymity for onion routing protocols \cite{Dingledine:2006:usability}.
They are characterized as complete network view due to a central authority, which distributes the list of Tor routers.
One exception is \cite{Mittal:2011:PIR-Tor}, which realizes private node retrieval and thereby constrains the decision maker's view of the network.
A complete view helps against adversary biasing node selection and is preferred in source-routing in order to prevent the decision maker to choose from a smaller set of nodes.\\
Further inherent routing features concerning the communication model include routing type, scheduling, determinism in the node selection, and the selection set. 
The exceptions here are~\cite{Tang:2010:Tor-CircuitScheduling, AlSabah:2012:DiffTor}, where they suggest a prioritization at the scheduling phase in favor of interactive traffic in order to reduce delays that interactive users might experience.


Node selection in all onion routing-based protocols is non-deterministic. 
This is important since the Tor network consists of volunteers and it is very likely to have a fraction of malicious nodes among them. 
A non-deterministic node selection reduces the chances of consistently selecting malicious nodes. 
Since the adversary is assumed to be local, a non-deterministic node selection makes targeted surveillance harder.\\ 
Furthermore, the node selection probability is generally weighted using static parameters, except for a few approaches that dynamically adjust weights, \eg
for balancing security versus performance \cite{Snader:2011:Tor-tuneup} and for avoiding congestion \cite{Wang:2012:Tor-CongestionAware,AlSabah:2012:Conflux}.
Onion routing protocols are low-latency and have circuit-based communication mode, which are both inherent routing features of these protocols.
Although we classify Tor as a protocol where the routing decision maker has a complete view, it is worth mentioning that the unlisted relays, known as \emph{bridges}, are not part of this view.

Next, we discuss random walk protocols and DHT-based protocols. 
Crowds are Morphmix are two of the early random walk protocols that were proposed for anonymous communication.
However, they present conceptual differences in terms of routing features.
Both \emph{Crowds} and \emph{Morphmix} have fully connected topologies since every node may build a connection with every other node, resulting in better availability of the system, which leads to a bigger attack surface for timing attacks. 

The path length in Crowds may vary and is determined in a non-deterministic manner to make simple timing attacks harder for external, local, and passive adversaries. 
Still, this does not necessarily hold for the case that at least one of the nodes in the path is malicious.
In Morphmix, the initiator does not select the nodes of the route herself, rather decides on the number of nodes and establishes the connection.

Crowds is semi-decentralized because routing information of nodes is distributed by a central entity (the blender), which introduces a single point of failure with respect to node administration.
Morphmix, however, has a fully decentralized structure.
The network view is complete in Crowds, which, on the one hand, protects Crowds from eclipse attacks and on the other hand, is important since Crowds has a hop-by-hop routing type that makes the node selection sensitive to be biased by adversaries.
In Morphmix, the network view is partial, and therefore, witnesses were introduced to prevent the biased node selection.
Moreover, an inherent feature of random walk protocols is that the node selection is non-deterministic. 
In Crowds, each node is chosen from the set of all nodes based on a geometric distribution~\cite{Danezis:2009:WisdomCrowds}; whereas, in Morphmix, the initiator knows a subset of nodes. 

An inherent routing feature of DHT-based protocols is partially connected topology and a partial network view. 
The routing information is distributed among nodes and no single node has the complete list. 
Such a design increases the scalability of the protocols.
A partially connected network topology makes DHT-based protocols less resilient against DoS attacks, which aim at disconnecting the network as much as possible compared to onion routing protocols.
The connection direction is bidirectional for the majority of protocols with two exceptions.
The exceptions are the file sharing applications Gnunet and Freenet Opennet mode. 

Generally, DHT-based protocols are fully peer-to-peer protocols. 
There are two exceptions in this category: namely, Torsk and Salsa, where the first one has a hybrid role structure while the latter one allows both hybrid and fully peer-to-peer role structures.
For being partially connected, DHT-based protocols provide a partial view of the network to the routing decision maker.
Note that this may introduce a series of attacks. 
Examples of attacks against protocols that provide only a partial view of the network to the routing decision maker are route fingerprinting attacks \cite{Danezis:2006:Fingerprinting}, and route bridging attacks \cite{Danezis:2008:RouteBridgingAttacks}.
Another series of attacks, which might be possible due to a partial network view, are attacks that aim at disconnecting target nodes from the rest of the network, such as eclipse attacks~\cite{Castro:2002:Eclipse}.

Most of the DHT-based protocols are characterized with a hop-by-hop routing type. 
Exceptions are NISAN, Salsa, and Octopus, with source-routing.
In Octopus, there are two decision makers for node selection;
the path initiator who decides only about a segment of the path and the last node of that segment, which initiates the rest of the path.
In our study, we could not find much information about the scheduling of DHT-based protocols, in particular for protocols that have not been deployed.
Most of the DHT-based protocols have non-deterministic node selection, again here exceptions are the file sharing applications, where the routing path does not need to be anonymous. 

The set selection for DHT-based protocols is, in most cases, all nodes within the routing table (\ie all nodes available to the decision maker). 
However, there are two exceptions: Torsk, where the set selection is restricted by security and network restrictions, and Freenet in the Darknet mode, where the set selection is based on trust assumptions of the user.
For most of DHT-based protocols, the selection probability is uniform, exceptions are Freenet and Gnunet.
Both protocols do not aim at providing unlinkability~\cite{Pfitzmann:2000:terminology} nor they hide that a user is participating in the network.
Nevertheless, they hide the role of the peer in the network.
%
Most of the DHT-based protocols are message-based except Torsk, AP3, and Salsa. 

Next, we discuss the DCNets protocols. 
\emph{DCNet-based protocols}, as classified in Table~\ref{table:onionrouting}, have some general inherent routing features that are due to the broadcast nature of their communication.
These inherent routing features include unidirectional connection, asynchronous connection, and network structure involving centralized entities.
Moreover, the routing type is source-routing with deterministic node selection and statically weighted selection probability. \\
Furthermore, DCnet-based protocols incur high latency and have message-based communication models.
In DCnets, we regard avoiding collusion and shuffling as routing relevant aspects of these protocols. 
The first designs of DCnets~\cite{Chaum:1988:DCNets, Waidner:1990:DC,Golle:2004:DC} and Dissent\cite{Corrigan-Gibbs:2010:Dissent} are direct realization of the original DCNet; therefore, they are similarly characterized.
Inherent characteristics of such protocols are fully connected network structures, having a fully peer-to-peer role model.
They support flat topologies, selecting all nodes for the selection set and offer a uniform selection probability for node selection.

In order to improve efficiency and performance, some DCNet-based protocols \cite{Wolinsky:2012:Dissent, Wolinsky:2012:Dissent-anytrust, Goel:2003:Herbivore} 
have been proposed, which vary in their routing features.
Unlike the first group, in these protocols, the network structure is partially connected.
For example, in Herbivore, participants are organized in star topologies, which are then connected in a ring topology. 
The organization of the nodes yields a hierarchical structure for the second group of DCnet protocols.
Moreover, in the extended version of Dissent, users do not share keys with each other but rather with designated servers. 
Furthermore, the new versions of DCnet-based protocols enforce network restrictions to the selection set in order to increase efficiency and performance.

We conclude this part of the discussion with miscellaneous protocols. 
\emph{Tarzan} protocol originally had a partially connected topology that was due to its partial network view of the route initiator. 
However, in the later version of Tarzan, a gossip-based strategy has been proposed to have a complete view for the route initiator, which leads to a fully connected topology as marked in Table~\ref{table:onionrouting}.
%
The connectivity of \emph{I2P} is similar to onion routing protocols due to the similarities for the node selection. 
I2P is characterized with unidirectional connection, which reduces the timing data that a single relay can have.
However, multiple relays participate in the communication between a sender and receiver.
The routing information of I2P is managed in a DHT-like fashion.
Each database node (floodfill peer) has a slice of the information \cite{Schimmer:2009:I2P}, which could enable adversaries to carry out eclipse attacks targeting floodfill nodes~\cite{Egger:2013:I2P-Practical-attacks}.
Since a user obtains node information from more than one floodfill node (up to eight), the union of this information might cover most of the I2P network and give the decision maker an almost complete view.
I2P uses a source-routing approach, allowing the users to choose nodes that are faster. 
The selection probability in I2P is non-deterministic with a bias towards nodes that are profiled as fast responding nodes. 
Response times of these nodes differ among users; hence, timing attacks are more difficult to mount compared to Tor, where the node selection is biased using publicly known information \cite{I2P:2016:peer-profiling}.
Since response times are continuously measured, we have marked the selection probability with a bias based on dynamic restrictions.
At the node level, I2P nodes use a prioritized scheduling mechanism, where each task has ``bid'', and the task with the lowest (best) bid is served first \cite{I2P:2016:transport}.


\subsection{Correlation, Conflicts, Trade-offs, and Applications}

In this section, we address correlation (\ie dependencies and conflicts), and trade-offs between routing characteristics of AC networks.
First, we review direct and indirect correlations of routing features by comparing them with each other. 
We conclude this section with a discussion about the relevance of specific routing characteristics for certain applications. 

\begin{table*}[htp]
\centering
\caption{Overview of the adversary definitions, focus of routing feature, and challenges that our four routing classes face}
\label{table:overviewdiscussion}
\begin{tabular}{l|l|l|l}
\hline
\textbf{Routing Class} & \textbf{Adversary Type} & \textbf{Routing Feature in Focus} & \textbf{Challenges} \\
\hline
\emph{Mixnet}& Global \& active &Forwarding (scheduling) \& node management (topology)& Traffic analysis attacks, such as flooding attacks \\

\emph{Onion routing} &Local \& active & Node selection &Traffic analysis attacks, \ie timing attacks \\

\emph{Random Walks (DHT)} & Local \& active &Node selection \& transfer of routing information& Partitioning attacks \& biasing node selection \\
 
\emph{DCnet} & Global \& passive &Forwarding&Collision and disruption\\
\hline
\end{tabular}
\end{table*}


We have defined the topology only based on connectivity of relaying routers (see Section~\ref{subsec:routing}).
This is the reason why users and administrative entities are not taken into account.
Hence, based on our definition, there is no correlation between topology and the two of the routing features, namely roles and decentralization.

There is an evident correlation between hierarchy and topology of AC networks.
A hierarchical AC network does not have a fully connected network structure. 
For example, Herbivore, which has a hierarchical routing strategy, has a partially connected topology.
Moreover, the network view of the routing decision maker can have an influence on the topology of the AC network. Generally speaking, a partial network view might lead to a partially connected network topology for the AC network because the routing decision maker might have difficulties accessing routing information of certain nodes. 
This holds for random walk and DHT-based protocols.
One exception is demonstrated by PIR-TOR, which uses PIR to keep the network view minimal, albeit the topology is fully connected. 
Therefore, the correlation between topology and the network view depends on further factors. 
For example, if the topology is partially connected, it might be that the routing decision maker has a partial view, but it also might be due to some other routing restrictions.

We also observe a correlation between topology and selection set. 
Namely, restrictions in the selection set lead to reduced connectivity of the network topology.
For example, in Restricted Route mix networks, the network view is complete; however, connectivity is restricted due to restrictions in the selection set, which leads to a partially connected network topology.


Although the synchronization of connections is not directly correlated to scheduling, it depends on the forwarding strategy of the particular nodes.
As mentioned in Section \ref{sec:mixnets}, the flushing algorithms influence synchronization when timed mixes are used.


AC networks with a hierarchical structure have partially connected network structure (\ie Herbivore and the extended version of Dissent).
By definition, hierarchical organization of nodes restricts the selection set.

Node management is more challenging in fully decentralized AC networks. Therefore, obtaining a complete view and a periodic updating of routing information is more difficult.
When the network view of the routing decision maker is partial, often source-routing has the advantage to prevent the bias of malicious nodes and partitioning attacks. 
Thus, AC protocols that use this combination need to employ a secure node selection policy in their protocol.
Examples of such protocols are Octopus and Morphmix. Octopus uses bound checking and proactively identifies malicious nodes; while, the latter one randomly selects witnesses to prevent bias in node selection.
A partial network view also restricts the selection set because the routing decision maker can select only nodes that it is aware of. 

Clearly, flushing algorithms also influence scheduling. 
For example, pool mixes can be defined to induce prioritized scheduling.
There is also a correlation between scheduling and latency because in a prioritized scheduling algorithm, some type of traffic is delayed.

Flushing algorithms also influence latency.
Timed mixes by themselves do not necessarily influence latency.
However, they might induce latency if long time restrictions have been selected. Same with the threshold mixes, when the incoming traffic is low compared to the threshold that has been set up.
There is also a correlation between latency and communication mode.
High latency AC networks usually use a message based communication mode and vice versa. 
This is because connections are not going to be used further (\eg replies are not going to be sent in a short time); therefore, setting up a circuit is unnecessary.

Next, we compare our four main groups by discussing their applications.
Mixnets are designed to be secure against traffic analysis and global adversaries by aggregating messages into batches. 
However, they are vulnerable to the collusion of mixnets and flooding attacks \cite{Serjantov:2003:Attacks}, in case if there are not enough (honest) users. 
Moreover, Mixnets' resilience against traffic analysis comes with a price and makes them more appropriate for high latency applications, such as emails and electronic voting.

Onion routing protocols, such as Tor, are more efficient (in particular faster) and have little computational overhead, making them suitable for low-latency applications, such as web browsing. 
Tor also leverages a large number of volunteer nodes. Almost all of these nodes are known to the routing decision maker.
However, the complete network structure for the routing decision maker can limit scalability.
Moreover, Tor is considered to be only secure against local adversaries and it is vulnerable to traffic analysis attacks \cite{Murdoch:2005:Traffic-Analysis, Chakravarty:2010:TA-BW, Evans:2009:Tor-Congestion-Attacks, Johnson:2013:trafficcorrelation, Mittal:2011:Tor-Attack, OGorman:2009:correlationattacks}, in particular if the adversary can access both ends of the communication.

Random walk protocols and protocols using DHT are designed rather for fully peer-to-peer networks, where the network view is incomplete. 
Having a fully peer-to-peer network motivates the growth of the network and helps scalability.
Therefore, they are suitable, for instance, for anonymous file sharing, where the nodes have to dedicate a considerable amount of resources. 
However, being fully peer-to-peer is considered to affect the usability of the protocol.
Unfortunately, this might lead to a decrease in the number of users of such systems and in turn reduce anonymity.
Last but not least, classic DCnets provide information-theoretic anonymity but some of them require a restricted setting, where all users or nodes need to be honest.
The classic DCnets were also not resilient against DoS attacks.
Moreover, DCnets tend do have a large communication overhead and do not scale well. 
Even Dissent, which employs a client-server approach for better scalability, can only scale up to a few thousand clients \cite{Wolinsky:2012:Dissent}. 
Therefore, they are more suitable for applications, such as micro-blogging, but at a small scale.

In Table \ref{table:overviewdiscussion}, we summarized the type of adversaries defined for the particular routing class. 
Moreover, the table shows the routing feature in focus for the particular class. 
Finally, the table lists challenges that our four routing classes face in terms of routing features and achieving anonymity and security.

\section{Concluding Remarks}
\label{sec:conclusion}
In this work, we classified anonymous routing characteristics. 
We identified main criteria groups, each with several routing features and dimensions tackling various aspects routing decisions in AC protocols. 
Moreover, we shortly described and then carefully evaluated the bulk of existing AC protocols under our classification. 
Furthermore, we discussed the relevance between routing decisions that are made in such networks and their influence on anonymity and security.
We have learned several lessons from conducting our survey. 
On the one hand, security, anonymity, scalability, and performance goals that are favored for anonymous communication are very hard to reach altogether, simply because the routing decisions, which support each of these goals, often contradict each other. 
This is especially true for achieving strong anonymity and good performance, which is still an open problem.
On the other hand, routing aspects are related to each other, for example, a partial view of the system (in the routing information) often supports the hop-by-hop routing. 
Therefore, it is very hard to separate the various routing aspects from one to another protocol. 
We observe that making certain routing decisions leads often to a trade-off between security, anonymity, scalability, and performance goals. 
Finally, our classification uncovers which routing decisions have to be tailored to the security, anonymity, scalability, and performance goals that are necessary for a specific use case of a given AC protocol.



\section*{Acknowledgment}
This work was supported in part by the Research Council KU Leuven: GOA TENSE (GOA/11/007) and the Flemish Government FWO G.0360.11N Location Privacy, FWO G.068611N Data mining and by the European Commission through H2020- DS-2014-653497 PANORAMIX and H2020-ICT-2014-644371 WITDOM. 
This work was also supported by Microsoft Research through its PhD Scholarship Programme. 
Moreover, this work was supported by the German Federal Ministry of Education and Research (BMBF) through funding for the Center for IT-Security, Privacy and Accountability (CISPA) (FKZ: 16KIS0345) and by the European Research Council Synergy Grant imPACT (n. 610150).


\bibliographystyle{ieeetr}

\bibliography{bib/references}

\end{document}